\documentclass[aps,shownopacs,superscriptaddress,nofootinbib,preprint,11pt]{revtex4}
 \pdfoutput=1
 \usepackage{hyperref}
\usepackage{amsmath}
\usepackage{float}
\usepackage{graphics}
\usepackage{graphicx}
\usepackage{epsfig}
\usepackage{psfrag}
\usepackage{color}
\usepackage{slashed}

\usepackage{amsfonts}
\usepackage{amssymb}

\newcommand*{\be}{\begin{equation}}
\newcommand*{\ee}{\end{equation}}
\newcommand*{\bea}{\begin{eqnarray}}
\newcommand*{\eea}{\end{eqnarray}}

\newcommand{\comment}[1]{}

\newcommand{\cref}[1]{Chapter~\ref{c.#1}}

\def\beq{\begin{equation}}
\def\eeq{\end{equation}}
\def\bea{\begin{eqnarray}}
\def\eea{\end{eqnarray}}
\def\ba{\begin{array}}
\def\ea{\end{array}}
\def\bi{\begin{itemize}}
\def\ei{\end{itemize}}
\def\be{\begin{enumerate}}
\def\ee{\end{enumerate}}
\def\bc{\begin{center}}
\def\ec{\end{center}}
\def\bt{\begin{table}}
\def\et{\end{table}}
\def\btb{\begin{tabular}}
\def\etb{\end{tabular}}

\def\lsim{\raise0.3ex\hbox{$\;<$\kern-0.75em\raise-1.1ex\hbox{$\sim\;$}}}
\def\gsim{\raise0.3ex\hbox{$\;>$\kern-0.75em\raise-1.1ex\hbox{$\sim\;$}}}

\begin{document}

\title{Revisiting neutrino masses from Planck scale operators}
\author{Abhishek M Iyer}
\email{abhishek@cts.iisc.ernet.in}
\affiliation{Centre for High Energy Physics, Indian Institute of Science,
Bangalore 560012}

\begin{abstract}
Planck scale lepton number violation is an interesting and natural possibility to explain non-zero neutrino masses. We consider such operators in the context of Randall-Sundrum (RS1) scenarios.
Implementation of this scenario with a single Higgs localized on the IR brane (standard RS1) is not phenomenologically viable as they lead to inconsistencies in the charged lepton
mass fits.
In this work we propose a set-up with two Higgs doublets. 
We present a detailed numerical analysis of the fits to fermion masses and mixing angles.
This model solves the issues regarding the fermion mass fits but solutions with consistent electroweak symmetry breaking are highly fine tuned.
A simple resolution is to consider supersymmetry in the bulk and a detailed discussion of which is provided. Constraints from flavour are found to be strong and Minimal Flavour Violation (MFV)
is imposed to alleviate them.
\end{abstract}
\vskip .5 true cm

\maketitle
\section{Introduction}
One of the most puzzling features of the Standard Model is the existence of extremely small but non-zero neutrino masses \cite{Tortola:2012te}.
Extension of the SM fermion content by the introduction of right handed neutrinos offers a possible solution to understand this intriguing aspect about neutrinos. 
Majorana mass terms for the right handed neutrinos are included in the lagrangian. If the Majorana mass is far greater that the Dirac mass \textit{i.e.} $M_R\gg m_D$
and assuming one flavour for simplicity, the light neutrino mass is given by the expression $m_\nu\sim\frac{m_D^2}{M_R}$.
This phenomenon is referred to as the `see-saw' mechanism \cite{Mohapatra}.
For $m_D\sim \mathcal{M_{EWSB}}\sim 200$ GeV, choice of $M_R \sim 10^{14}$ GeV will result in light neutrino mass $m_\nu\sim 0.01$ eV.
This is called see-saw mechanism of Type-1 where the right handed fermion is a singlet under the $SU(2)_L$ gauge group.
There are two other types of see-saw mechanism referred to as Type-II and Type-III. In Type-II, a scalar triplet is added to the SM matter content.
Type-III also contains a right handed fermion but is a triplet under $SU(2)_L$.
 A detailed review of the various types of see-saw mechanism
can be found for example in \cite{Strumia:2006db,Davidson:2008bu}. 

The see-saw mechanisms ascribe a Majorana nature to the neutrino which violate lepton number. 
It is not yet known whether neutrinos are of Dirac or Majorana type.
Though there is no experimental evidence favoring either of them, theoretically the realization of Dirac type neutrino seems less natural than Majorana neutrinos as it 
requires the assumption of global lepton number conservation. 
This is because the assumption of conservation of global symmetries lead to problems in fundamental theories of quantum gravity \footnote{
It has been shown that Dirac neutrinos are possible even in the presence of explicit lepton number violating terms in the RS model, where lepton 
number violation can be hidden in the extra-dimensions\cite{planckgher,Iyer:2013hca}}\cite{Witten}

It is also possible to generate small neutrino masses without extending the SM field content. This is possible by the introduction of higher dimensional lepton
number violating operators of the form $LLHH$. These operators are in general suppressed by the Planck scale. This is because the origin of this operator
can be traced up to the lepton number breaking effects at the quantum gravity scale which is typically the Planck scale. In 4D however, such operators with $\mathcal{O}$(1) co-efficients
lead to neutrino masses of $\mathcal{O}\left(\frac{v^2}{M_{Pl}}\right)$ which are very small. They 
would require an enhancement in the co-efficients of these operators of the $\mathcal{O}(10^3-10^4)$ to generate neutrino masses near the atmospheric scale.
This is not viable as it would lead to non-perturbativity in the calculations. Instead such operators suppressed by sub-Planckian scales ($\sim 10^{14}$ GeV) are then introduced to generate the neutrino masses. However, in addition to the introduction
of an intermediate scale (between the Planck and EW scale), it also weakens the motivation of their Planckian scale origins.

In an attempt to revive the Planck scale suppressed operators, we shall consider them in the context of extradimensional
models.  One advantage of considering such models is that in the effective 4D theory the suppressing scale is in general less than the Planck scale. This is because the effective 4D scale is defined
in terms of the Planck scale as $\Lambda=f_{bulk}M_{Pl}$, where $f_{bulk}$ is a function of fundamental bulk parameters of the theory.
A particular realization of the extra-dimensional framework we consider here, is the one proposed by Randall and Sundrum \cite{RS}. It consists of a single extra-dimension compactified on an $S_1/Z_2$
orbifold. A 3-brane is introduced at each of the orbifold fixed points \textit{i.e.} at $y=0$ and $y=\pi R$. The presence of a large negative bulk energy density attributes
a warped geometry to the bulk. Introduction of brane-localized sources results in a vanishing cosmological constant on the branes. Identifying the scale
of physics at the $y=0$ brane as the Planck scale, the effective UV scale induced at the brane at $y=\pi R$ is given as $e^{-kR\pi}k$ where $R$ is the radius of compactification
and $k$ is the reduced Planck scale. Choosing $kR\sim 11$ we find that Planck scale masses are naturally warped down to the $TeV$ scale on the $y=\pi R$ brane
with $\mathcal{O}$(1) choice of model parameters thus providing an elegant solution to the hierarchy problem. The geometry of RS also offers a natural explanation to the observed hierarchical masses of fermions by means of the split fermion approach introduced in \cite{ArkaniHamed} and
applied to the RS framework in \cite{Hisano}.

Neutrino masses and flavour phenomenology in RS model have been previously considered in \cite{gross,Kitano,Huber2,Huber3,Huber4,Agashe1,Chen,AgasheSundrum,Fitzpatrick,a,b,c,d}.
More recently in \cite{Iyer} a detailed analysis of various models of neutrino mass generation was considered.
In particular, the Planck scale lepton number violating operators was studied. It was observed that in a regular RS setup, realization of Planck scale lepton number violating operators (LLHH)
led to unattractive solutions as far as fitting of charged lepton masses are concerned. 
In this case the small neutrino mass was determined entirely by the bulk mass parameters for the leptonic doublets. In scenarios with large warping ($\sim 10^{-16}$), the 
zero mode for the doublets was
localized towards the UV brane to fit $eV$ scale neutrino masses. To offset the UV localization of the zero mode doublets, the charged singlets are required
 to be localized very close to the IR, near the Higgs, to fit the charged lepton masses. This leads to non-perturbative Yukawa couplings between the zero mode singlets and the higher KK modes. 
 For flavour processes this rendered the higher order diagrams incalculable. In addition
 this scenario is not tenable as it leads to reintroduction of large hierarchies among the bulk mass parameters.

In this work we revisit the LHLH operator in a regular RS setup. 
The goal of the study is to revive the LLHH type scenario wherein it
could lead to a consistent description of lepton mass and mixing data. To our knowledge this is the first explicit neutrino mass model in a regular RS setup (with warp factor 
$\epsilon \sim 10^{-16} $) wherein Planck scale operators is consistent phenomenologically.
 We consider a two Higgs doublet model of Type II. One Higgs doublet ( say $H_u$) couples to the neutrino (up-sector)
 while the other Higgs doublet (say $H_d$) which couples to the charged lepton sector (down sector).
 The key idea is to disentangle the Yukawa couplings for the up and the down sector. The presence of two bulk Higgs presents itself with following four possibilities
 for the localization of the zero mode:\newline A) The zero mode of $H_u$ is localized close to the IR brane while that of $H_d$ is localized close to the UV brane. \newline
B)The zero mode of $H_d$ is localized close to the IR brane while the zero mode of $H_u$ is localized close to the UV brane.
The localization of the zero modes are flipped with respect to configuration A. However, $H_u$ cannot be localized very close to the UV brane as it will lead to very small neutrino masses. \newline
C)The zero mode of both the Higgs doublets are localized near the UV brane.
Similar constraints on the localization of $H_u$ applies in this case as well. If $H_u$ and $H_d$ are localized at the same point near the UV brane then this scenario is equivalent
to a single bulk Higgs scenario. \newline
D)The zero mode of both the Higgs doublets are localized near the IR brane.
$H_d$ cannot be strongly localized near the IR brane as it will lead to the re-introduction of hierarchies in the bulk masses of the charged singlets as observed in \cite{Iyer}.

All the configurations results in a situation where the lepton masses and mixing angles can
  be fit with $\mathcal{O}(1)$ choice of bulk mass parameters. Technique of $\chi^2$ minimization introduced in \cite{Iyer} is adapted to identify
 the parameter space of bulk masses which fit the data. Configurations A and B are less favoured than C and D when the hadronic sector fits are taken into account.
Electro-weak symmetry breaking for all the four configurations can only be realised with large fine tuning in the $\mathcal{O}$(1) couplings of the two Higgs doublet potential.
This renders some of the couplings to be non-perturbative.
We propose a solution by super-symmetrizing the model.
 Constraints from lepton flavour and particularly the loop induced decays are very strong.
 The contribution to such decays is very large for low lying KK masses of fermions.
 Supersymmetric contributions are neglected by raising the soft mass scale.
 The ansatz of Minimal Flavour Violation (MFV) is then applied to alleviate the bounds. It is to be noted that while the solution to the hierarchy
 problem may be compromised in this case owing to the particular localization of Higgs fields, the assumption of supersymmetry helps resolve the issue.
 
 The rest of the paper is organized as follows: We begin with a brief discussion of the LLHH operator in RS model with a single Higgs in Section[\ref{section2}]. This provides 
 the necessary  motivation to consider a set-up with an additional Higgs in the bulk. Yukawa coupling of fermions with a bulk Higgs are derived in Section[\ref{section3}]. 
 Numerical fits to the lepton mass and mixing angles are discussed. We also present the results of the fits to the hadronic sector.
 In Section[\ref{section4}] we address a few issues regarding the Electroweak symmetry Breaking (EWSB) specific to the scenario under consideration.
 We find that significant amount of fine tuning is required for all four configurations to satisfy the electroweak symmetry breaking conditions.
The scenarios can be rescued by imposing 4D supersymmtery in the presence of bulk multiplets and is discussed in Section[\ref{section4a}].
 Lepton Flavour violation is
 considered in Section[\ref{section5}]. The details regarding the coupling of fermions to the scalar mass eigenstates are presented in the Appendix[\ref{couplings}]. In Section[\ref{section6}]
 Minimal Flavour Violation and it's application to leptonic case with no right handed neutrinos is considered. We conclude in Section[\ref{section7}].
 
 \section{Planck scale lepton number violating operators in standard RS scenarios}
 \label{section2}
 In this section providing the necessary motivation to consider models with two Higgs doublets in the presence of Planck scale lepton number violating operators.
 Consider a single Higgs doublet ($H$) localized on the IR brane. The Yukawa part of the action for the leptons is given as
\begin{equation}
 S_{Yuk}\supset \int d^4x dy \delta(y-\pi R) \left[\left(\kappa\frac{1}{\Lambda^{(5)}} LLHH+Y_E\bar LHE_R\ldots\right)\right]
\label{actionyuk1}
 \end{equation}
where $\Lambda^{(5)}\sim 2.2\times 10^{18}$ GeV. Here $L(E)$ stands for doublet(singlet).
  Using the following KK expansion for the bulk fields
 \begin{eqnarray}
\label{kkexpansion}
L_{L} (x, y) = \sum_{n=0}^\infty \frac{1}{\sqrt{\pi R} } e^{2 \sigma(y) } L_{L}^{(n)}(x) f^{(n)}_{L }(y) &;&L_{R}(x, y)= \sum_{n=0}^\infty {1 \over \sqrt{\pi R} } e^{2 \sigma(y) } L_{R}^{(n)}(x) \chi^{(n)}_{L }(y) \nonumber \\
E_{R}(x,y)= \sum_{n=0}^\infty \frac{1}{\sqrt{\pi R} } e^{2 \sigma(y) } E_{R}^{(n)}(x) f^{(n)}_{E }(y)&;& E_{l}(x,y)= \sum_{n=0}^\infty {1 \over \sqrt{\pi R} } e^{2 \sigma(y) } E_{L}^{(n)}(x) \chi^{(n)}_{E}(y)\nonumber\\
\end{eqnarray} 
where $L,R$ denote chirality. We assume $L_L,E_R$ to be even under $Z_2$ and hence non-vanishing on the brane while $L_R,E_L$ are odd. 
Integrating over $y$, we arrive at the following effective 4D mass matrix for the neutrino
\begin{eqnarray}
 (m_{\nu})_{ij}&=&\kappa'_{ij}\mathcal{N}_{c_{L_i}}\mathcal{N}_{c_{L_j}}\epsilon^{c_{L_i}-c_{L_j}-2}\nonumber\\
 &\sim&\frac{\kappa'_{ij}}{\epsilon \Lambda^{(5)}}\epsilon^{c_{L_i}-c_{L_j}-1}\;\;\;\;\;\;\text{$c_L>0.5$}
\label{n1}
 \end{eqnarray}
 where $\kappa'=2k\kappa$ and $\epsilon =e^{-kR\pi}\sim10^{-16}$.
 The normalization factor is given as
 \begin{equation}
 \mathcal{N}_{c_a}=\sqrt{\frac{0.5-c_a}{\epsilon^{2c_a-1}-1}}
 \label{normalizationfactor}
\end{equation}
where $a=L,R$. Since $\epsilon \Lambda^{(5)}\sim \text{TeV}$, we find that $c_L\sim 0.9$ is required to fit neutrino mass $\mathcal{O}$(0.04) eV.
  Turning our attention to the charged leptons, the mass matrix can be obtained from Eq.(\ref{actionyuk1}) as 
\begin{equation}
 (m_{E})_{ij}=\mathcal{N}_{c_{L_i}}\mathcal{N}_{c_{E_j}}\epsilon^{c_{L_i}-c_{E_j}-1}
\label{e1}
 \end{equation}
 In order to fit the small neutrino masses, the zero mode of the doublets are required to be very close to the UV brane ($c_L\sim 0.9$).
 To offset the UV localization of the doublets, the corresponding zero modes for the charged singlets are required to be localized very
 close to the IR brane to fit the charged lepton masses. Quantitatively, the $c$ values for the charged singlets vary from $-100$ to $-10^{7}$ from the first to the third generation,
 thus resulting in the reintroduction of hierarchies in the bulk mass parameters. The bulk masses are above the cutoff scale and the effective 4D Yukawa coupling of the
 the SM singlets to the KK modes are non-perturbative.
 As a result a setup with a Higgs localized near the IR brane is not very attractive. 
 
 \section{Planck scale lepton number violation with bulk Higgses}
 \label{section3}
 We now explore Planck scale lepton number violating operators in a RS set up with bulk Higgs. 
 In \cite{Iyer:2013eka}, such an operator were considered in a  RS model with a smaller warp factor ($\sim 10^{-2}$) and was shown to generate correct neutrino masses.
 In this case however, we consider the RS model with large warp factor $\sim 10^{-16}$ as in the original setup \cite{RS}. 
 An RS model with two Higgs doublets,
 labeled as $H_u$ and $H_d$ is studied.  
  $H_u$ couples to the up(neutrino) sector
 while $H_d$ couples only to the down(charged lepton) sector. Such models are referred to as two Higgs doublet models of Type-II.
 The basic idea is to decouple the neutral lepton and the charged lepton Yukawa couplings.
 The Yukawa part of action for the leptonic and hadronic sector with two Higgs doublets is given as
  \begin{equation}
 S_{Yuk}\supset \int d^4x dy \left[\left(\kappa\frac{1}{\Lambda^{(5)}} LLH_uH_u+Y_E\bar LH_dE+Y_U\bar Q\tilde H_uU+Y_D\bar Q H_dD\right)\right]
\label{actionyuk}
 \end{equation}
 where $\tilde H_u=i\sigma_2H_u^*$.
Here $H_u$ and $H_d$ are the two Higgs doublets.
The presence of two Higgs fields in the bulk provides two additional parameters $b_u$ and $b_d$ for $H_u$ and $H_d$ respectively. We will show that $b_d$ and $b_u$ can be suitably manipulated to find a solution where the
lepton masses and mixing angles  can be fit by $\mathcal{O}$(1) choice of bulk mass parameters $c_E$ and $c_L$.
For a bulk scalar field in a warped background, the bulk zero mode solutions are not consistent with the boundary conditions (Neumann) \cite{gherghetta,Gherghetta1}.
Brane localized mass terms are added giving rise to modified Neumann boundary conditions with the brane mass
parameter appropriately adjusted to result in consistent zero mode solutions.
Bulk action for a complex scalar field $\Phi$ is given as \cite{gherghetta,Gherghetta1}
\begin{equation}
 S=\int d^4xdy\sqrt{-g}\left[\partial_M\Phi^*\partial^M\Phi+\left(m_\Phi^2+ 2bk\delta(y)-2bk\delta(y-\pi R)\right)|\Phi|^2\right]
 \label{freeaction}
 \end{equation}
where we parametrize the bulk mass as $m_\Phi^2=ak^2$ with $a,b$ being dimensionless quantities. Here $\Phi$ denotes $H_u$ or $H_d$ and correspondingly $b$ denotes $b_u$ or $b_d$.
Ideally one would expect them to be $\mathcal{O}$(1).
The bulk field $\Phi$ is KK expanded as
\begin{equation}
 \Phi(x,y)=\frac{1}{\sqrt{\pi R}}\sum_{n=0}\phi(x)f^{(n)}_\phi(y)
\end{equation}
The zero mode profile for a bulk scalar is given as \cite{gherghetta,Gherghetta1}
\begin{equation}
 f_\Phi^{(0)}(y)=\sqrt{k\pi R}\zeta_\Phi e^{bky}
 \label{scalarzero}
\end{equation}
where 
\begin{equation}
\zeta_\Phi=\sqrt{\frac{2(b-1)}{\epsilon^{2(1-b)}-1}}
\end{equation}
where $\epsilon=e^{-kR\pi}$ is the warp factor. 
The brane parameter $b$ must be tuned to be $b=2\pm\sqrt{4+a}$ to satisfy the boundary conditions for the zero modes.
To understand the localization property of the zero mode, consider the canonically normalized profile given as
$\tilde f_\Phi^{(0)}=\sqrt{k\pi R}\zeta_\Phi e^{(b-1)ky}$.
$b>1(b<1)$ implies the zero mode of the Higgs is localized towards the IR(UV) brane.

For a bulk Higgs, the fundamental Yukawa couplings $Y^{(5)}_E$ have mass dimension -1/2. 
Using the KK expansion in Eq.(\ref{kkexpansion}) and integrating
over the extra-dimension the zero mode mass matrix for all charged fermions in general is given as 
\begin{equation}
 m_{ij}=v_dY'_{ij}\zeta_\phi \mathcal{N}_{L_i} \mathcal{N}_{E_j}\left(\frac{\epsilon^{(c_{L_i}+c_{E_j}-b_d)} -1}{b_d-c_{L_i}-c_{E_j}}\right)
 \label{diracmassa}
\end{equation}
where we have defined the dimensionless $\mathcal{O}$(1) Yukawa coupling as $Y'_E=2\sqrt{k}Y^{(5)}_E$. The normalization factor $\mathcal{N}_{c_a}$ is given by Eq.(\ref{normalizationfactor}).

Similarly the  expression for the neutrino mass matrix can be determined from Eq.(\ref{actionyuk}) as
\begin{equation}
  (m_\nu)_{ij}=\frac{v_u^2}{2\Lambda}\kappa'_{ij}\mathcal{N}_{L_i} \mathcal{N}_{L_j} \zeta_{\Phi}\left(\frac{\epsilon^{(c_{L_i}+c_{L_j}-2b_u)} -1}{2b_u-c_{L_i}-c_{L_j}}\right)
\label{neutrinomassa}
\end{equation}
where $\kappa'=2\sqrt{k}\kappa$.
We adapt the technique of $\chi^2$ minimization introduced in \cite{Iyer} to identify the range of bulk paramters of the doublets and charged singlets for all the  four configurations which admit
a good fit to the data. The $\chi^2$ function is defined as
\begin{equation}
 \chi^2=\sum_{i=1}^N\frac{\left(O_i^{exp}-O_i^{theory}\right)^2}{\sigma_i^2}
 \label{chisqb}
\end{equation}
$O_i^{theory}$ is the model prediction for the value of the $i^{th}$ observable while $O_i^{exp}$ is its corresponding experimental number measured with a uncertainty $\sigma_i$.
The central values of the observables( $O_i^{exp}$) along with the corresponding $\sigma_i$ at the electroweak scale have been taken from \cite{pdg,valle}.
The fitting of charged lepton masses to high accuracy with which it is measured requires very precise tuning of the model parameters.
Thus we allow up-to $1.5\%$ errors in the masses of the charged leptons which gives a good estimate of the parameter space of model parameters which fit the data.
The ratio of the vev of $H_u$ and $H_d$ was chosen to be $tan\beta=10$.
All the parameters \textit{i.e.} the fifteen Yukawa couplings  and the six $c_{L,E}$ parameters are varied so as to minimize the function in Eq.(\ref{chisqb}). We consider the points 
which give a  $\chi^2$ between 0 and 10 to be a `good fit' to the data.
The $\mathcal{O}$(1) Yukawa couplings were varied between -10 and 10 with a minimum of 0.08 on the magnitude to avoid un-naturally small Yukawa paramters. All the bulk
mass parameters are varied between 0 and 1. We assume neutrino masses to have a normal hierarchy. Fits do not change considerably if inverted hierarchy is assumed.

We make the following choices for the bulk parameters of bulk Higgs fields and briefly summarize the results of the fit for all the cases:\newline
 (A) configuration: $b_u=3.0$ for $H_u$ and $b_d=0.3$ for $H_d$ corresponding
 to IR and UV localized scalar fields respectively. In this case the lepton doublets are localized near the UV brane. The presence of an additional Higgs doublets in the form
 of $H_d$ facilitates a fit to the charged lepton masses with $\mathcal{O}(1)$ parameters. As shown in Fig.[\ref{lhlhfita}] the singlets are localized near the IR brane
 with the corresponding $c$ values $>-0.5$. As can be expected, owing to it's larger mass, the third generation singlet is likely to be localized closer to $H_d$ as compared to the first two generation\newline
 (B) configuration: $H_u$ is UV localized and $H_d$ is IR localized. We choose $b_u=0.65$ and $b_d=3.0$. The results in this case will be nearly opposite to that of configuration A.
 As shown in Fig.[\ref{lhlhfitb}] the zero mode of the doublets are localized near the IR brane. The corresponding charged singlets are localized away from the zero mode of $H_d$
 near the UV brane.\newline
 (C) configuration: $b_u=b_d=0.65$ corresponding to both the zero modes localized near the UV brane.\newline
 Note that for configuration B and C the zero mode $H_u$ cannot be sharply localized towards the UV brane ($b_u<0.5$) as it would lead to very small neutrino masses.
 In this case $b_d=0.65$ was chosen form simplicity and $b_d>0.65$ is also permitted. The doublets and the first two generation singlets are predominantly localized near the IR brane as shown in Fig.[\ref{lhlhfitc}.].
 \newline
 (D) configuration: Both the zero modes are localized near the IR brane corresponding to the choice $b_u=3.0$ and $b_d=1.01$.
 In this case the zero mode of $H_d$ cannot be sharply localized towards the IR brane as it will lead to similar issues faced with a single Higgs localized on the IR brane discussed
 in Section[\ref{section2}]. As can be seen from Fig.[\ref{lhlhfitd}] This case is quantitatively similar to Configuration A as the changing $b_d$ from 0.3 to 1.01 does not significantly change the results of the fit. 
 
 The range
 of bulk masses $c_{L,E}$ for all the four configurations which satisfy the minimum $\chi^2$ requirement are summarized  in Table \ref{ranges}. 

 \begin{table}
\begin{tabular}{|cccc||cccc|}
 \hline
 \multicolumn{4}{|c||}{A}&\multicolumn{4}{|c|}{B}\\
 parameter & range &parameter & range&parameter & range &parameter & range\\
 \hline
 $c_{L_1}$ & 0.81-1.0 &$c_{E_1}$ &0.16-0.38&$c_{L_1}$ & 0.0-0.46 &$c_{E_1}$ &0.66-0.99 \\
$c_{L_2}$ & 0.81-0.99 &$c_{E_2} $&0.32-0.42&$c_{L_2}$ & 0.0-0.43 &$c_{E_2} $&0.62-0.99 \\
$c_{L_3}$ & 0.82-0.97 &$c_{E_3} $&0.40-0.80& $c_{L_3}$ & 0.12-0.56 &$c_{E_3} $&0.48-0.99  \\ 
\hline
 \multicolumn{4}{|c||}{C}&\multicolumn{4}{|c|}{D}\\
$c_{L_1}$ & 0.08-0.99 &$c_{E_1}$ &0.06-0.65&$c_{L_1}$ & 0.87-1.0 &$c_{E_1}$ &0.13-0.32 \\
$c_{L_2}$ & 0.32-0.63 &$c_{E_2} $&0.32-0.99&$c_{L_2}$ & 0.83-0.96 &$c_{E_2} $&0.34-0.41 \\
$c_{L_3}$ & 0.41-0.61 &$c_{E_3} $&0.41-1.0&$c_{L_3}$ & 0.83-0.95 &$c_{E_3} $&0.44-0.99  \\
\hline
\end{tabular}
\caption{Summary of ranges for the bulk mass parameters for all the four configurations.}
\label{ranges}
\end{table}

The reader might be inquisitive whether such localization of the two Higgses in the bulk would lead to correct fits for the hadronic sector or not.
The hadronic fits for configurations
A and B is challenging as far the third generation is concerned. For these two configurations, the zero modes of the two Higgs doublets are localized towards each of the orbifold
fixed point (one towards IR and the other towards the UV).
This leads to a tension in the fitting of the bottom and top quark masses. The large top mass requires the third generation doublet to 
be closer to the localization of $H_u$ thereby rendering the bottom quark mass fit to be small with $\mathcal{O}$(1) Yukawa parameters. 
This problem can be mildly alleviated by choosing
the $c$ value for the third generation doublet to be $c_{Q_3}\sim 0.5$. This corresponds to a delocalized third generation doublet.
With this choice it is possible to fit both the bottom and the top quark mass by choosing the Yukawa parameters for the third generation
in the range between 11 and 12. Allowing the third generation doublet to be localized either brane will require the $\mathcal{O}$(1) Yukawa coupling for either the top or the
bottom mass to be much larger than 12. This represents a slightly fine tuned case.

Configuration C and D on the other hand are more favoured since they allow fits for both the leptonic and the hadronic sector with choice of $\mathcal{O}$(1) Yukawa between 0.1 and 10. 
The $c$ parameters for the light quarks for both the configurations are scanned between 0 and 1. $c_{Q_3}$ and $c_{U_3}$ for configuration C is scanned between 0.5 and 3. For configuration
D they are scanned between -3 to 0.5. This is owing to the large top quark mass which favour the localization of the third generation doublet and the top singlet closer to the zero mode of 
$H_u$. We present the results of the scan for the quark sector for only these configurations and are given in Table[\ref{rangesquark1}] and \ref{rangesquark2}.
 \begin{table}
\begin{tabular}{|cccccc|}
 \hline
 
 parameter & range &parameter & range& parameter &range\\
 \hline
 $c_{Q_1}$ &0.24-0.99&$c_{D_1}$&0.2-1.0&$c_{U_1}$& 0-1.0\\
$c_{Q_2} $&0.33-1.0&$c_{D_2}$&0.3-1.0&$c_{U_2}$&0.3-1.0 \\
$c_{Q_3}$&0.63-3.0&$c_{D_3}$&0.44-1.0&$c_{U_3}$&0.6-3.0  \\ 
\hline
\hline
\end{tabular}
\caption{Summary of ranges for the bulk mass parameters for configuration C for the hadronic sector.}
\label{rangesquark1}
\end{table}

 \begin{table}
\begin{tabular}{|cccccc|}
 \hline
 parameter & range &parameter & range& parameter&range\\
 \hline
 $c_{Q_1}$ &0.2-0.91&$c_{D_1}$&0-1&$c_{U_1}$& 0-0.72\\
$c_{Q_2} $&0.36-0.64&$c_{D_2}$&0-1&$c_{U_2}$&0-0.62 \\
$c_{Q_3}$&-3.0-0.49&$c_{D_3}$&0-0.35&$c_{U_3}$&-3-0.47  \\ 
\hline

\end{tabular}
\caption{Summary of ranges for the bulk mass parameters for configuration for D hadronic sector.}
\label{rangesquark2}
\end{table}

\begin{figure}[htp]
\begin{tabular}{c}
\includegraphics[width=0.9\textwidth,angle=0]{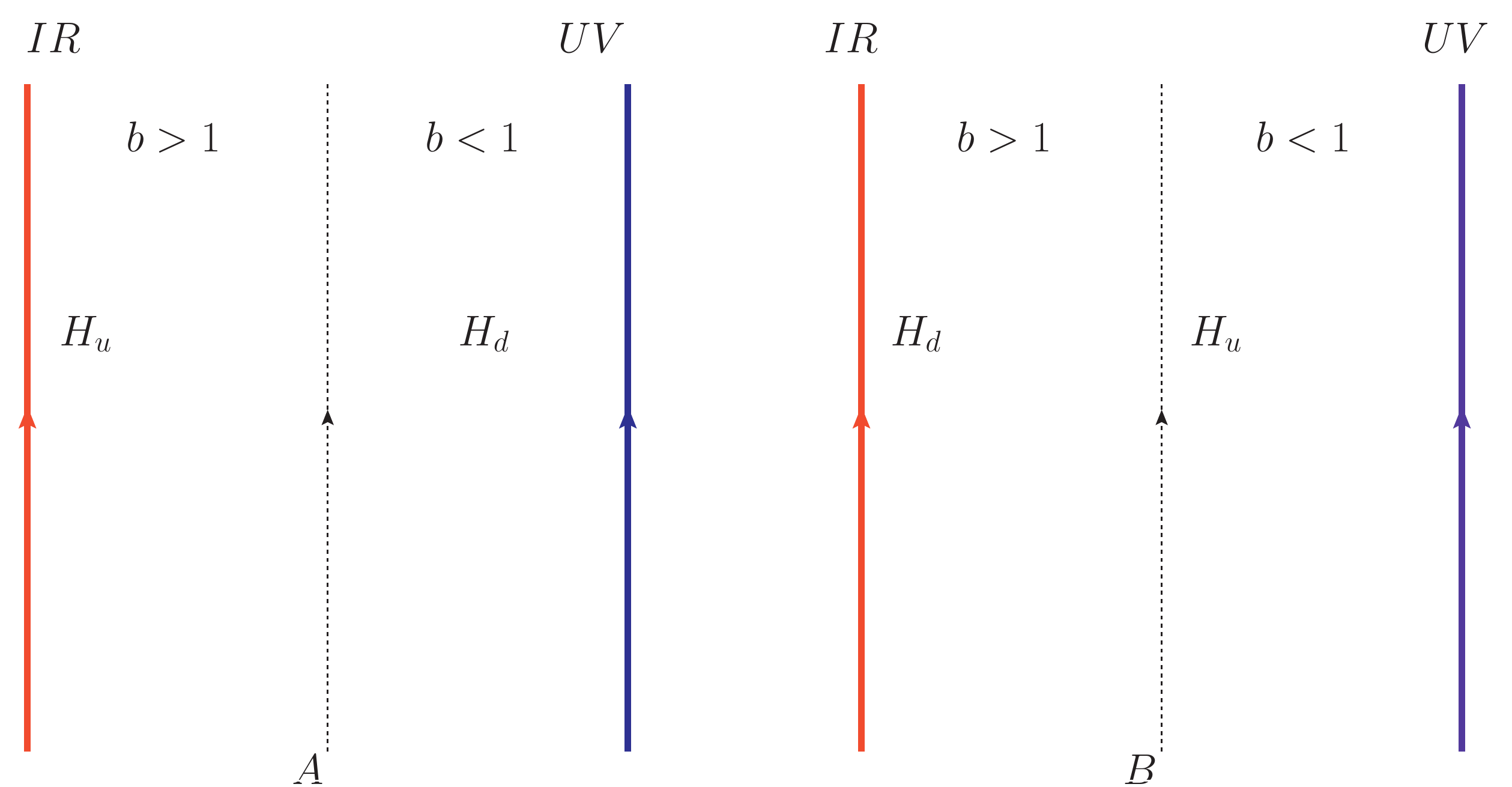} \\
\includegraphics[width=0.9\textwidth,angle=0]{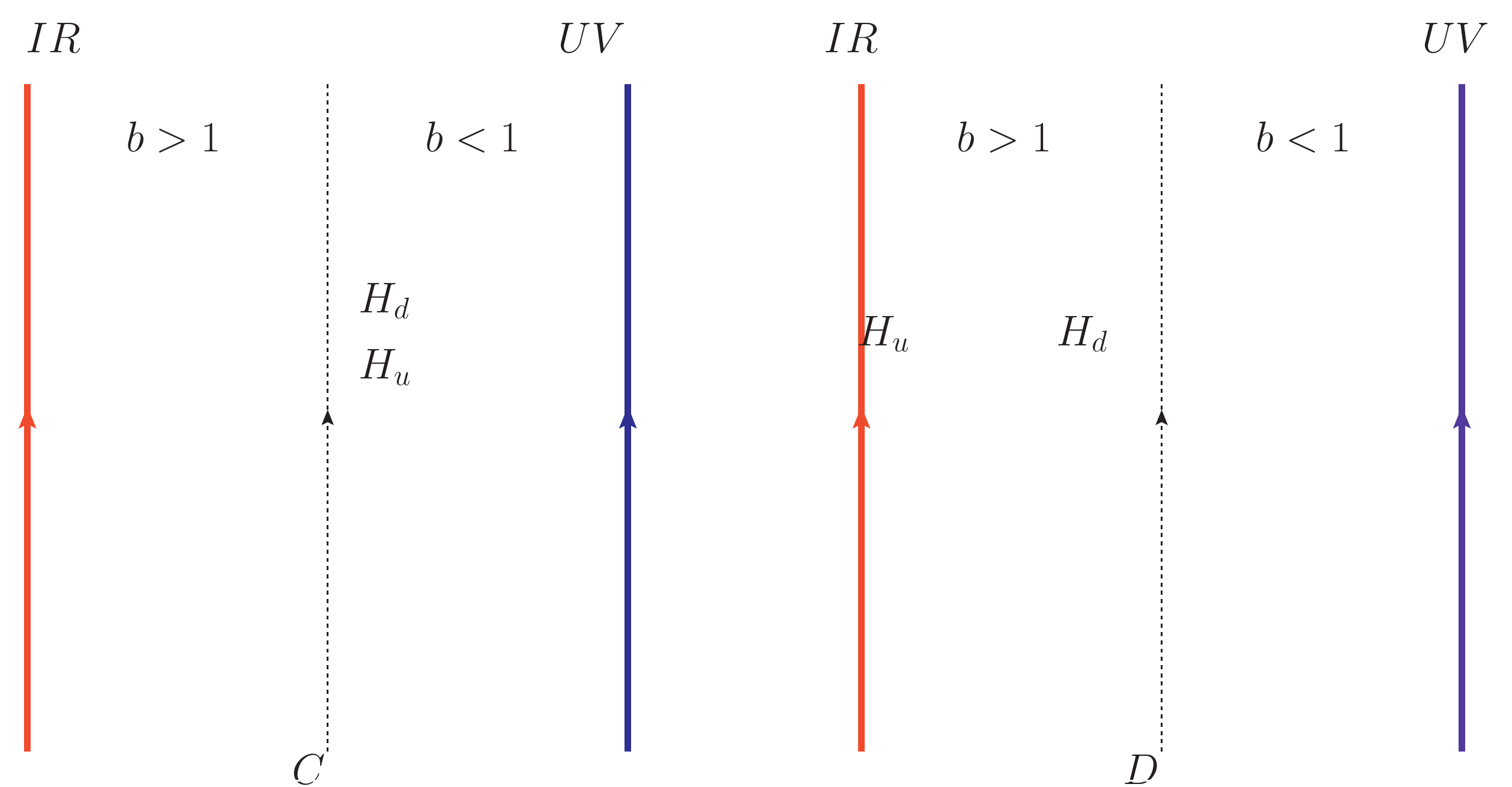} 
\end{tabular}
 \caption{The four different configuration which depict the localization of the zero mode of the Higgs doublet in the bulk. }
\label{configuration}
\end{figure}

\section{Issues with non-supersymmetric models}
\label{section4}
As we have seen in Section[\ref{section3}], the brane mass parameters ($b$) could be appropriately adjusted to arrive at the four configurations for the localization
of the zero mode\footnote{In principle the presence of $m^2_{12}$ in Eq.(\ref{potential}) can modify the profiles. We will assume it to be small compared to $m^2_{u}$ and $m^2_{d}$. We can choose $m_{12}\sim 0.1 k$}. However the zero modes for the both the Higgs fields were massless. It would be interesting to see it's implications on the vacuum conditions of the
effective zero mode potential. In this section we quantitatively analyze the conditions for electroweak minima for each of the four configurations.

We begin with the following two Higgs doublet potential with bulk masses $m^2_{u,d}$ and $m^2_{12}$.
\begin{eqnarray}
 V^{(5)}&=&m^2_{u,d}|H_{u,d}|^2-m^{2}_{12}\left(H^\dagger_dH_u+H^\dagger_uH_d\right)+\frac{\lambda_1}{2}\left(H_d^\dagger H_d\right)^2+\frac{\lambda_2}{2}\left(H_u^\dagger H_u\right)^2\nonumber\\
 &+&\lambda_3H_d^\dagger H_dH_u^\dagger H_u + \lambda_4 H_d^\dagger H_uH_u^\dagger H_d+\frac{\lambda_5}{2}\left[\left(H_d^\dagger H_u\right)^2+\left(H_u^\dagger H_d\right)^2\right]
\label{potential}
 \end{eqnarray}
 KK expanding the fields $H_{u,d}$, 4D effective potential for the zero mode is given as 
  \begin{eqnarray}
 V^{(4)}&=&-m^2_{12}H^{(0)}_uH^{(0)}_dK^{(0,0)}_{ud}+\frac{\lambda'_1}{2}(H^{\dagger(0)}_dH^{(0)}_d)^2G^{(0,0)}_{dd}+\frac{\lambda'_2}{2}(H^{\dagger(0)}_uH^{(0)}_u)^2G^{(0,0)}_{uu}+V_{cross}
\label{potentiala}
 \end{eqnarray}
 where $\lambda'_i=k\lambda_i$ are dimensionless parameters. 
 The cross terms in the potential $V_{cross}$ is given as
 \begin{equation}
  V_{cross}=\left[\lambda'_3(H^{\dagger(0)}_dH^{(0)}_d)^2(H^{\dagger(0)}_uH^{(0)}_u)^2+\lambda'_4(H^{\dagger(0)}_dH^{(0)}_u)(H^{\dagger(0)}_uH^{(0)}_d)+\frac{\lambda'_5}{2}\left((H^{\dagger(0)}_dH^{(0)}_u)+u\leftrightarrow d\right)\right]G^{(0,0)}_{ud}\\
 \end{equation}

 The dimensionless overlap integrals $K,G$ are defined as
\begin{eqnarray}
 K^{(m,n)}_{ab}=\frac{1}{\pi R}\int dy~\sqrt{-g}f^{(m)}_{H_a}f^{(n)}_{H_b} \;\;\;\; G^{(m,n)}_{ab}=\frac{1}{(\pi R)^2}\int dy~\sqrt{-g}(f^{(m)}_{H_a})^2(f^{(n)}_{H_b})^2
\label{overlap1}
 \end{eqnarray}
 $f^{(n)}_{H_a}$ denotes the  bulk profile for the $n^{th}$ KK state of $H_a$ where $a=1,2$.
 Note that the bulk mass term for the scalar included in the above potential are used to determine the bulk wave-functions of the field. 
 Thus in the two Higgs doublet terminology
 \cite{Aoki,Branco}, this results in $m^2_{1}=m^2_{2}=0$ at the effective 4D level.
  The correction to the potential due to mixing between the zero mode and the higher KK modes are neglected.
 We now make a few observations on the conditions required for electroweak symmetry breaking
 by investigating the potential at the minimum. 
We assume the potential is minimized along the following direction
\begin{eqnarray}
 H_u=\begin{pmatrix}
      0\\\frac{v_2}{\sqrt{2}}
     \end{pmatrix}\;\;\;\;\; H_d=\begin{pmatrix}
      0\\\frac{v_1}{\sqrt{2}}
     \end{pmatrix}
\end{eqnarray}
The effective 4D potential at it's minimum is given as
\begin{equation}
 V^{(4)}_{min}=-m^2_{12}v_1v_2K^{(0,0)}_{ud}+\frac{\lambda'_1}{2}\frac{v_1^4}{4}G^{(0,0)}_{dd}+\frac{\lambda'_2}{2}\frac{v_2^4}{4}G^{(0,0)}_{uu}+
 \left[\lambda'_3+\lambda'_4+\lambda'_5\right]\frac{v_1^2v_2^2}{4}G^{(0,0)}_{ud}
\label{potentialmin}
 \end{equation}
 The minimization conditions corresponding to the potential in Eq.(\ref{potentialmin}) \textit{i.e.} $\frac{\partial V^{(4)}_{min} }{\partial v_1}=0$ and $\frac{\partial V^{(4)}_{min} }{\partial v_2}=0$,
can be easily derived to be given as
\begin{eqnarray}
\frac{\lambda'_1}{2}v_1^2G^{(0,0)}_{dd}=m^2_{12}\frac{v_2}{v_1}K^{(0,0)}_{ud}-\left[\lambda'_3+\lambda'_4+\lambda'_5\right]\frac{v_2^2}{2}G^{(0,0)}_{ud}\nonumber\\
\frac{\lambda'_2}{2}v_2^2G^{(0,0)}_{uu}=m^2_{12}\frac{v_1}{v_2}K^{(0,0)}_{ud}-\left[\lambda'_3+\lambda'_4+\lambda'_5\right]\frac{v_1^2}{2}G^{(0,0)}_{ud}
\label{minimizationcondition}
\end{eqnarray}
 For a moderate $\tan\beta$, $v_2\sim 240$ GeV and $v_1 = \frac{v_2}{tan\beta}$. Under this assumption the left hand side and right hand side of both equations of Eq.(\ref{minimizationcondition}) must have the same
 order of magnitude (numerically consistent). We briefly consider the implications of each configuration of bulk Higgses on EWSB.\newline
  (A) configuration: $H_u$ is localized near the IR brane and $H_d$ is localized near the UV brane: The overlap integral $G^{(0,0)}_{ab}$ is negligible since the zero modes are localized away from
  each other.
 As a result the integral in Eq.(\ref{minimizationcondition})
 can be simplified to 
 \begin{eqnarray}
\frac{\lambda'_1}{2}v_1^2G^{(0,0)}_{dd}=m^2_{12}\frac{v_2}{v_1}K^{(0,0)}_{ud}\nonumber\\
\frac{\lambda'_2}{2}v_2^2G^{(0,0)}_{uu}=m^2_{12}\frac{v_1}{v_2}K^{(0,0)}_{ud}
\label{minimizationconditiona}
\end{eqnarray}
 Additionally since $m^2_{12}$ is a bulk mass
 is typically $\mathcal{O}$($k^2$). Warping down this scale to $M_{EWSB}^2$ would require the overlap integral $K^{(0,0)}_{ud}$ to be $\mathcal{O}(10^{-32})$ which
 is possible if $b_u\sim 3$.
Evaluating the overlap integrals, the conditions in Eq.(\ref{minimizationconditiona}) can be reduced to
\begin{equation}
(\tan\beta)^4\frac{b_u-1}{1-b_d}=\left(\frac{\lambda'_1}{\lambda'_2}\right)
 \label{mincondtion}
\end{equation}
Thus the choice of $\tan\beta\sim 5$, $b_u\sim 3$ and $b_d\sim 0.3$ would require the $\lambda'_1$ coupling to be unnaturally large for  $\lambda'_2 \sim \mathcal{O}$(1).\newline
(B) configuration: $H_u$ is localized near the UV brane and $H_d$ is localized near the IR brane: Conclusions similar to configuration A apply to this case as well with the exception that 
$b_d\sim 3$ is required to warp down the $m_{12}^2$ bulk mass to electroweak scale.\newline
(C) configuration:$H_u$ an $H_d$ are localized near the UV brane: The overlap integrals $G^{(0,0)}_{ab}$ in this case are $\mathcal{O}$(1). However, the $m^2_{12}$ term on the RHS of Eq.(\ref{minimizationcondition})
does not receive any suppression resulting in the the set of equations to be numerically inconsistent. \newline
(D) configuration: $H_u$ and $H_d$ are localized near the IR brane: As noted in Section[\ref{section3}], $H_d$ cannot be localized very close to the IR brane as issue of non-perturbative coupling
and hierarchical $c_E$ values would reappear. As a result $b_u\sim 3$ and $b_d\sim1$ are chosen for the fits. After evaluating the integrals in Eq.(\ref{minimizationcondition})
for $b_d,b_u>1$ we get
\begin{eqnarray}
\frac{\lambda'_1}{2}v_1^2(b_d-1)&=&h_1(b_d,b_u)\tilde m^2\frac{v_2}{v_1}+h_2(b_d,b_u)\left[\lambda'_3+\lambda'_4+\lambda'_5\right]\frac{v_2^2}{2}\nonumber\\
\frac{\lambda'_2}{2}v_2^2(b_u-1)&=&h_1(b_d,b_u)\tilde m^2\frac{v_1}{v_2}+h_2(b_d,b_u)\left[\lambda'_3+\lambda'_4+\lambda'_5\right]\frac{v_1^2}{2}
\label{minimizationconditionb}
\end{eqnarray}
where $\tilde m^2=e^{-2kR\pi}m^2_{12}$ and $h_1,h_2$ are $\mathcal{O}$(1) functions of $b_d,b_u$ defined as
\begin{eqnarray}
h_1(b_d,b_u)=\frac{2\sqrt{(b_d-1)(b_u-1)}}{b_d+b_u-4}\nonumber\\
 h_2(b_d,b_u)=\frac{2(b_d-1)(b_u-1)}{b_d+b_u-2}
\end{eqnarray}
Since $m_{12}\lesssim k$ and for the given choices of $b_d,b_u$,
the warp factor in the first term on the RHS of Eq.(\ref{minimizationconditionb}) ensures that it is $\lesssim \mathcal{O}(M_{electroweak})$.
Numerically, choosing $m_{12}\sim 0.1 k$ yields $\tilde m_{12}\sim 50$ GeV. Additionally, for $b_u= 3.01$ and $b_d=1.01$ we find that $h_1=14$ and $h_2=0.01$.
The $\lambda'$ couplings are also constrained from the requirement that the potential in Eq.(\ref{potentiala}) is bounded from below. Explicitly, it is given as \cite{Aoki,Branco}
\begin{eqnarray}
 \lambda'_1\geq 0\;\;\;\;\; \lambda'_2\geq 0\;\;\;\;\;\;  \lambda''_3\geq -\sqrt{\lambda'_1\lambda'_2}\;\;\;\;\;  \lambda''_3+\lambda''_4-|\lambda''_5|\geq -\sqrt{\lambda'_1\lambda'_2}
\label{boundedness}
 \end{eqnarray}
where $\lambda''_i=h_2(b_d,b_u)\lambda'_i$ for $i=3,4,5$. 
A consistent set of solutions for $\lambda'$ and $\lambda''$ couplings which simultaneously satisfy both Eq.(\ref{minimizationconditionb}) and Eq.(\ref{boundedness}) requires a significant amount of fine tuning.
For instance choosing  $tan\beta=10$ and $\tilde m_{12}=50$ GeV we get $\lambda_2'\sim 10^{-5}$ and $\lambda''_4\sim 11$ for $\mathcal{O}$(1) choices of the other parameters
\footnote{For a given $tan\beta$ and $\tilde m_{12}$, $\lambda''_4$ and $\lambda'_2$ are fairly independent of the choices of the other $\lambda$ parameters up-to $\mathcal{O}$(1)
 variations.}. This implies that $\lambda''_4$ and therefore the $\lambda'_4$ (since $h_2\sim 0.01$) coupling is non-perturbative.
Without resorting to a full numerical fit, we conclude that the case of pure SM with 2HDM in RS requires significant fine tuning in the $\lambda'$ couplings.
We now proceed to look for alternative solutions.
\section{Supersymmetric extensions}
\label{section4a}
In the previous section we saw that all the four configurations of Higgses in the bulk had issues with electroweak symmetry breaking. 
A possible solution to this problem is to supersymmetrize the model. Supersymmetry breaking introduces soft terms \footnote{We will assume SUSY breaking on the brane.} of the form $\tilde m^2_{i}$ which could be helpful in alleviating
the fine tuning necessary to achieve EWSB. In addition, $m^2_{12}$ which are also generated due to soft breaking effects will not contribute to the scalar profiles in the bulk.
In this section we discuss the implications of supersymmetrizing the model for all the four configurations.

The extension to supersymmetric scenarios is fairly straightforward. 
Thus far we have stressed on two Higgs doublet of Type II in which one Higgs doublet ($H_u$) couples to the up sector (neutrinos) while the other doublet ($H_d$) couples to the down sector 
(charged leptons) fermions. This construction is exactly similar to the Yukawa part of the superpotential where two Higgs doublets are necessary for the superpotential to be holomorphic.
To see the relation between the mass formula between the supersymmetric and non-supersymmetric case, consider bulk hypermultiplets of the form $\Phi_Z=(Z,Z^c)$. $Z$ and $Z^c$ are $N=1$ chiral multiplets and $Z=L,H_{u,d},E,U,D$.
This results in $N=2$ supersymmetry at the effective 4D level.
We assume $Z$ and $Z^c$ to have opposite $Z_2$ parities such that at the zero mode level we are left with purely $N=1$ chiral multiplets.
The 5D superpotential with bulk fields can then be written as \cite{choi1,choi2}
\begin{equation}
 \mathcal{W}_{YUK}=e^{-3\sigma(y)}\left(Y_ULH_uU+Y_ELH_dE+Y_DLH_dD+\frac{\kappa}{\Lambda^{(5)}}LLH_uH_u+\ldots\right)
\label{superpotential}
 \end{equation}
where $L,U,D,H_{u,d}$ are $N=1$ chiral multiplets and $\sigma(y)=k|y|$. The dots in Eq.(\ref{superpotential}) represents terms of the form $LH_uE^c$ which gives rise to corrections to the fermion
masses at the loop level.
The bulk profile for the zero mode of the chiral multiplet $X$ is given as \cite{Marti}
\begin{equation}
f^{(0)}_Y=e^{\frac{3}{2}-c_Y}
\label{susybulk}
\end{equation}
 where $c_Y$ represents the bulk
mass of the corresponding hypermultiplet $\Phi_Y$. Using Eq.(\ref{susybulk}) in Eq.(\ref{superpotential}) and working in the basis in which the K{\"a}hler terms are canonically normalized
we find that the expressions for the masses are exactly the same as in Eq.(\ref{diracmassa}) and (\ref{neutrinomassa}). The parameter $b_{u,d}$ is identified with the bulk
mass of the corresponding hypermultiplet as $b_{u,d}=\frac{3}{2}-c_{H_u,H_d}$. As a result, our analysis using two Higgs doublet model of Type II can be directly applied
to supersymmetric extensions\footnote{For configurations A and B the fits must be done with $b_u\sim 1$ and $b_d\sim 1$ respectively. This is to prevent the warping of 
$m_{12}$ generated due to soft supersymmetry breaking effects.}.

\subsection{Soft breaking terms}
Supersymmetry breaking terms are generated on the brane by the interaction of the bulk fields with the supersymmetry breaking spurion $X$. It is parameterized as $X=\theta^2 F$.
The spurion $X$ is chosen to be localized on the brane near which $H_u$ is localized. This is necessary to generate large $X_t$ to have $m_{Higgs}=125$ GeV with light stop masses.
\begin{equation}
\mathcal{K}^{(4)}=\int dy \delta(y-a)e^{-2 k \pi R}k^{-2}X^\dagger X\left(\beta_{Y,ij}Y^\dagger_iY_j\right)
\end{equation}
where  $\beta$ have dimensional carrying negative mass dimensions of -1 ( as the matter fields are five dimensional).  

The soft masses are generated when the $X$ fields get a vacuum expectation value. 
The sfermion mass matrix will however not be diagonal in flavour space.  In the canonical basis the mass matrices are of the form \cite{choi1,choi2,Dudas,nomura,Iyer2a}
\begin{equation}
 (m_{\tilde{f}}^2)_{ij} = m_{3/2}^2~\hat \beta_{ij}~e^{(1-c_i-c_j)k R \pi}\mathcal{N}_i\mathcal{N}_j
\label{softmassmatrix2}
\end{equation} 
where $\hat \beta_{ij}=2k \beta_{ij}$ are dimensionless $\mathcal{O}(1)$ parameters. $\mathcal{N}_i$ are defined in Eq.(\ref{normalizationfactor}). The gravitino
mass is defined as 
\begin{equation}
m_{3/2}^2 = { <F>^2 \over k^2 } = { <F>^2 \over M_{Pl}^2 }
\end{equation}

The localization of the superfields at different points in the bulk also lends a flavourful structure to the soft mass matrices. For $X$ localized at the $y=a$
orbifold fixed point, they are given as
\begin{equation}
 \tilde m^2_{ij}=m^2_0 \zeta_\phi \mathcal{N}_{i} \mathcal{N}_{j}e^{(1-c_i-c_j)ka}
\label{softterms}
 \end{equation}
where  $m_0$ is the soft mass scale.
The presence of SUSY breaking effects is useful on many accounts:
First the mass term $m^2_{12}$ are generated only after supersymmetry is broken. As a result it helps in justifying
their non-inclusion to determine the bulk profiles of the scalar fields. Secondly, the choice of $b_u=3$ in configuration A and $b_d$ in configuration B is necessary to warp 
down $m^2_{12}$ (which is $\mathcal{O}(M_{PL})$) to the electroweak scale. Since these terms are generated due to supersymmetry breaking effects they are 
naturally of $\mathcal{O}$(TeV) or below.

\section{Lepton Flavour Violation}
\label{section5}
 Mixing of the SM fermions with the KK states of various fields gives rise to additional contributions to the flavour changing processes.
 The rare flavour violating decays can arise at tree level($l_j \to l_i l_k l_k$) and at the one loop diagrams of the form $l_j \to l_i + \gamma$.
 In the presence of SUSY, flavour violation in the presence of soft terms would contribute with the corresponding neutralinos/charginos. In the following we will assume $m_0$ to be $> 5$ TeV as a result of which the MSSM contributions to the flavour processes are suppressed. The contributions to the flavour processes
due to the KK states of the superpartners is similar to the MSSM contribution.
This can also be suppressed by assuming the lowest KK scale \textit{i.e} $k\epsilon\sim 5$ TeV \cite{Larsen:2012rq}. However we shall see that even by assuming a lowest KK scale as large as
5 TeV is not sufficient to suppress the FCNC due to the SM KK states in our model.\newline
 \textbf{Tree Level:} 
The contribution to tree level decays are predominantly due to the non-universal coupling of the zero mode fermions to the first KK state of the Z boson ($Z^{(1)}$).
In the presence of two Higgs doublets the physical Higgs spectrum consists of 2 CP even scalars, one pseudo scalar and two charged Higgs.
Sub leading contributions due to  tree level exchange of neutral scalars (h,H and A) also arise and depend on the mixing angle $\delta$ (defined in Equation(\ref{mixingangle})), 
which parametrizes the mixing between the zero mode and the higher KK modes of the scalars. As shown in Fig.[\ref{o1}], it is 
at-most $\mathcal{O}(1)$  depending on the position of localization of the Higgses. 
$\delta$ is negligible for configurations A and C, since zero mode of $H_d$ is localized close to the UV brane.
For configurations B and D where the zero mode of $H_d$ is localized near
the IR brane ($b>1$) implying $\delta$ is $\mathcal{O}(\frac{v^2}{M^2_{kk}})$. 
In addition, this contribution also depends on the coupling $Y_d^{001}$ (defined in Eq.(\ref{nmk}) of left and right chiral zero mode fermion
to the first KK state of $H_d$. As shown in Fig[\ref{lhlhfitb}] for configuration B, since the majority of points, especially for the first two generation
which fit the masses correspond to $c>0.5$, the overlap of the corresponding zero mode with the KK state of $H_d$ will be small.
Similarly for configuration D, the zero mode for the doublets are also localized very close to the UV brane as shown in [\ref{lhlhfitd}]
which will also result in the corresponding coupling $Y_d^{001}$ to be very small.
As a result we do not consider their effects
for tree level processes.

The tree level processes $j\rightarrow ikk$ can be parametrized by the following effective terms in the lagrangian
\begin{eqnarray}
-\mathcal{L_{{eff}}}&=&
\frac{4G_F}{\sqrt{2}}\left[\beta^{ij}_3(\bar{i}_R\gamma^\mu j_R)(\bar{k}_R\gamma_\mu k_R)
+\beta^{ij}_4(\bar{i}_L\gamma^\mu j_L)(\bar{k}_L\gamma_\mu k_L)\right. \nonumber \\
& &+\left.\beta^{ij}_5(\bar{i}_R\gamma^\mu j_R)(\bar{k}_L\gamma_\mu k_L)
+\beta^{ij}_6(\bar{i}_L\gamma^\mu j_L)(\bar{k}_R\gamma_\mu k_R)\right] + {\rm h.c.}
\label{ngeq}
\end{eqnarray}
where $i,j,k$ denote flavour indices and the flavour changing vertex is parameterized by $\beta^{ij}_\alpha$ ($\alpha=3,4,5,6$).
The expression for the branching fraction in terms of the co-efficients $\beta^{ij}_\alpha$ can be found in \cite{Agashe1}.
Fermion fields whose zero modes are localized close to the UV brane \textit{i.e.} $c>0.5$
or are localized close to the IR brane \textit{i.e.} $c<-5$, will 
couple universally to the first KK mode of the gauge boson \cite{Gherghetta1,Iyer}.
We now discuss the implications of fermion localization on tree-level FCNC for all the four configuration of Higgs in the bulk.

\begin{figure}[h]
  \begin{tabular}{cc}
  \includegraphics[width=0.50\textwidth,angle=0]{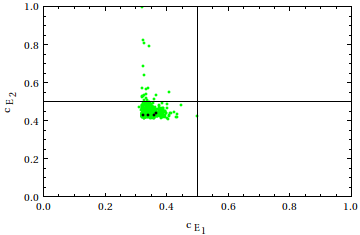}&
  \includegraphics[width=0.50\textwidth,angle=0]{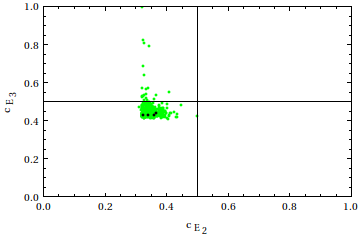}
  \end{tabular}
\caption{Green region depicts total available parameter space of $c_E$ which fit charged lepton masses for configuration A.
The black points represent the available parameter space 
after the constraints from tree level processes ($l_i \to l_j l_k l_k$) are imposed for $M_{KK}=2$ TeV.}
  \label{llhhavailable}
  \end{figure}

(A) configuration: $H_u$ is localized near the IR brane and $H_d$ is localized near the UV brane: As shown in Table[\ref{ranges}] and in Fig.[\ref{lhlhfita}], the neutrino mass fits require the lepton doublets close to the UV brane.
Thus the  doublets couple universally to $Z^{(1)}$, giving negligible contribution to tree level FCNC.
 The charged singlets, however have non-universal coupling to gauge KK states which could constrain the available parameter
 space for $c_E$. 
  Figure [\ref{llhhavailable}] depicts the total parameter space of the charged singlets in green and available parameter space in black after the constraints from tree
 level FCNC are imposed. The mass of the first KK state of $Z$ is chosen to be 2 TeV for the calculations. From the figure we see that there is a small amount parameter space
 (represented by black dots) available which satisfy the constraints from tree level processes.
 
 (B) configuration:  $H_u$ is localized near the UV brane and $H_d$ is localized near the IR brane:
 As shown in Table.[\ref{ranges}], since the doublets and the singlets (especially the first two generation) have a tendency to be localized near the UV brane, the constraints
 from tree level FCNC are weak. As a result it is possible to find points which satisfy the constraints for a lowest gauge KK mass of around $2$ TeV.
 
 (C) configuration: The configuration where both the Higgs are localized near the UV brane is far more constrained. As seen in Table.[\ref{ranges}], unlike the case B,
 the doublets no longer couple universally to the KK state of $Z$. Along with the singlets, they give rise to appreciable contribution to the tree level
 FCNC. Thus the minimum gauge KK scale required to satisfy the constraints from all tree level processes increases to around 11 TeV.

 (D) configuration: Both the Higgs doublets are localized near the IR brane. 
 Since the distribution of $c$ parameters in this case is very similar to configuration A the conclusions for this case are similar to configuration A. 

\begin{figure}[h]
  \begin{tabular}{c}
  \includegraphics[width=1.\textwidth,angle=0]{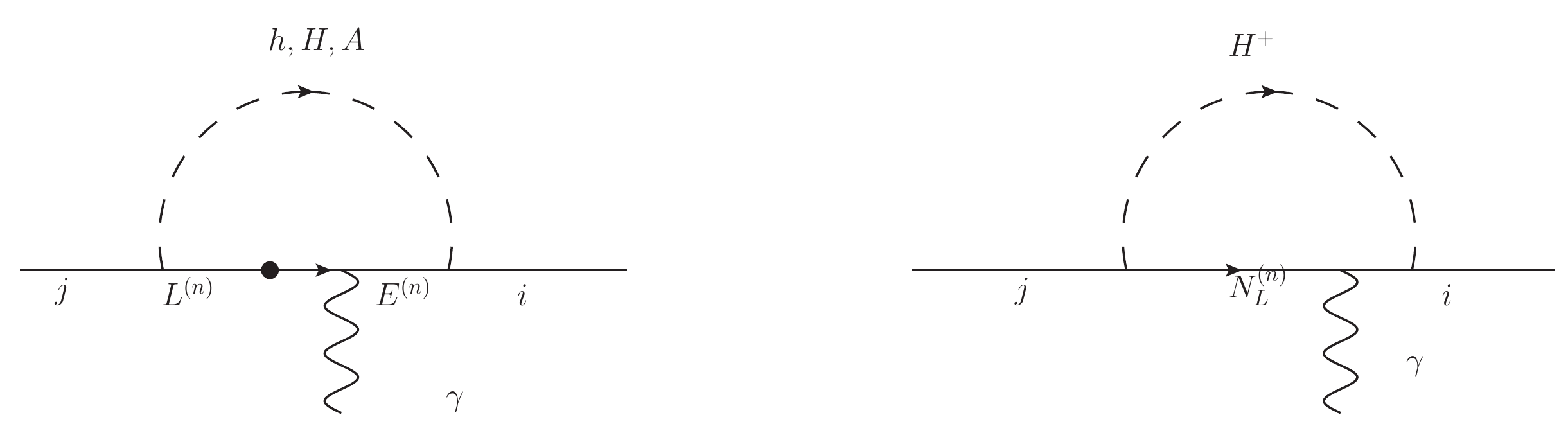}
  \end{tabular}
\caption{The figure shows the contribution to process $j\rightarrow i\gamma$ due to exchange of neutral scalar mass eigenstates ($h,H,A$) on the left and charged Higgs on the right.}
  \label{llhhdipole}
  \end{figure}

 \textbf{Loop level graphs:}  As shown in Fig.[\ref{llhhdipole}] in addition to the physical Higgs ($h$) in the loop,
  there will be additional contributions due to heavier CP even eigenstate ($H$), pseudo scalar ($A$) and the charged Higgs graph ($H^+$). The contribution due to the Goldstone boson
  in the $R_\xi$ gauge will be similar
  to the charged Higgs graph. The effective lagrangian for the process $j\rightarrow i \gamma$ can be parameterized as
  \begin{eqnarray}
-\mathcal{L_{{\rm eff}}}&=&
A_R(q^2)\frac{1}{2m_j}\bar{i}_R\sigma^{\mu\nu}F_{\mu\nu}j_L+
A_L(q^2)\frac{1}{2m_j}\bar{i}_L\sigma^{\mu\nu}F_{\mu\nu}j_R \nonumber \\
\label{ngeq1}
\end{eqnarray}
In terms of the co-efficients $A_L,A_R$ the Branching fraction for the loop induced process is given as
\begin{equation}
 BR(l_j\rightarrow l_i\gamma)=\frac{12\pi^2}{(G_Fm_j^2)^2}(A_L^2+A_R^2)
\label{BR}
 \end{equation}
 The discussion of loop induced decays for the four configurations of Higgses can be divided into two categories:\newline
1) $H_d$ localized near the UV brane.\newline
This case corresponds to configurations A and C discussed in Section[\ref{section2}].
For this case let us first consider the contribution due to the exchange of neutral scalar states in the loop as shown in the left panel of Fig.[\ref{llhhdipole}].
The mass-insertion in the internal KK fermion line (represented by a dot) can be expressed by the following integral
\begin{equation}
 g^{(n,m)}=\frac{1}{(\pi R)^{3/2}}\int dy f^{(n)}(y)f^{(m)}(y)f^{(0)}_{H_d}(y)
\end{equation}
where $f^{(0)}_{H_d}$ is defined in Eq.(\ref{scalarzero}) and $f^{(n)}(y)$ is the profile for the $n^{th}$ KK mode of the fermion and is given as \cite{Gherghetta1}
\begin{equation}
 f^{(n)}(y)=\frac{e^{ky}}{N_n}\left[J_\alpha(\frac{m_n}{k}e^{ky})+b_\alpha Y_\alpha(\frac{m_n}{k}e^{ky})\right]
\end{equation}
where $\alpha=\sqrt{\frac{1}{4}+c(c\pm1)}$ for the $L(R)$ fields and $N_n$ is the normalization factor. $J_\alpha(Y_\alpha)$ are Bessel's function of first(second) kind with order $\alpha$.
 \begin{figure}[h]
  \begin{tabular}{c}
  \includegraphics[width=.5\textwidth,angle=0]{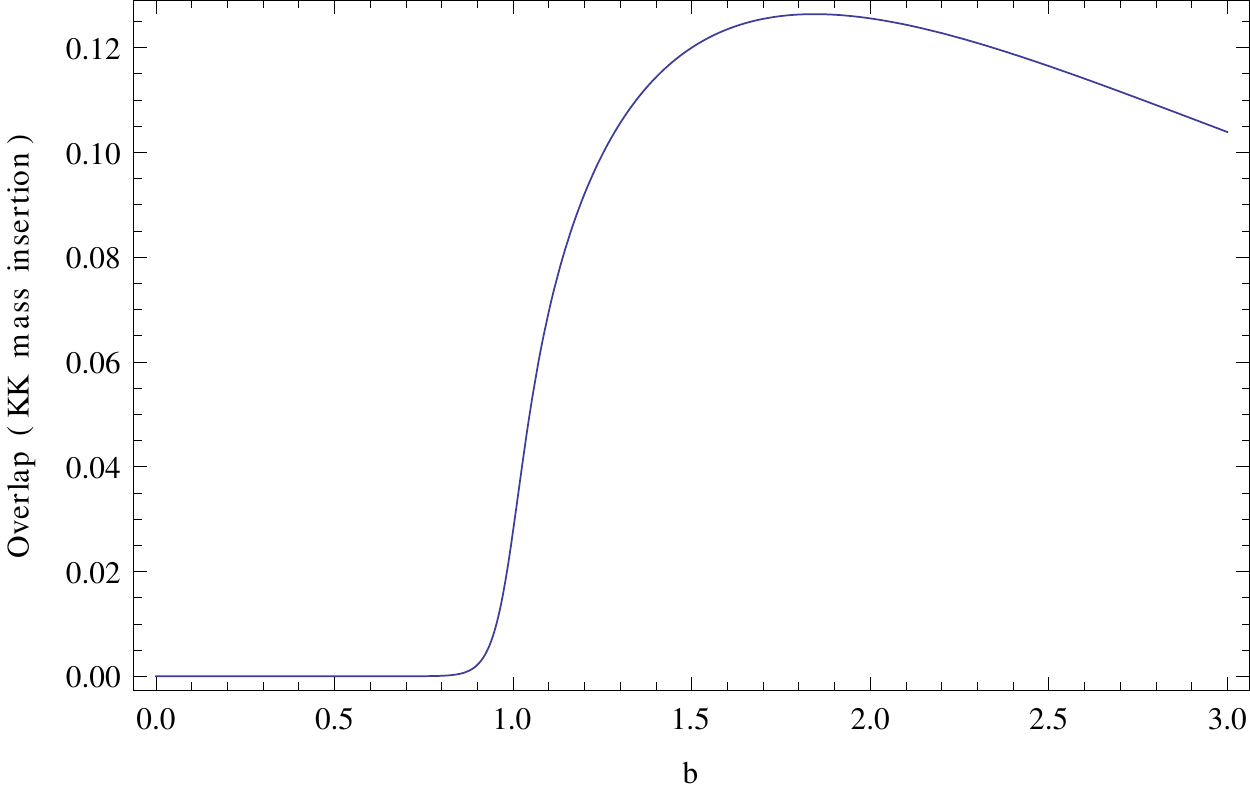}
  \end{tabular}
\caption{The figure shows the overlap between two fermion KK fields and the vev in the mass insertion approximation.}
  \label{o2}
  \end{figure}
  
\begin{figure}[h]
  \begin{tabular}{c}
  \includegraphics[width=.5\textwidth,angle=0]{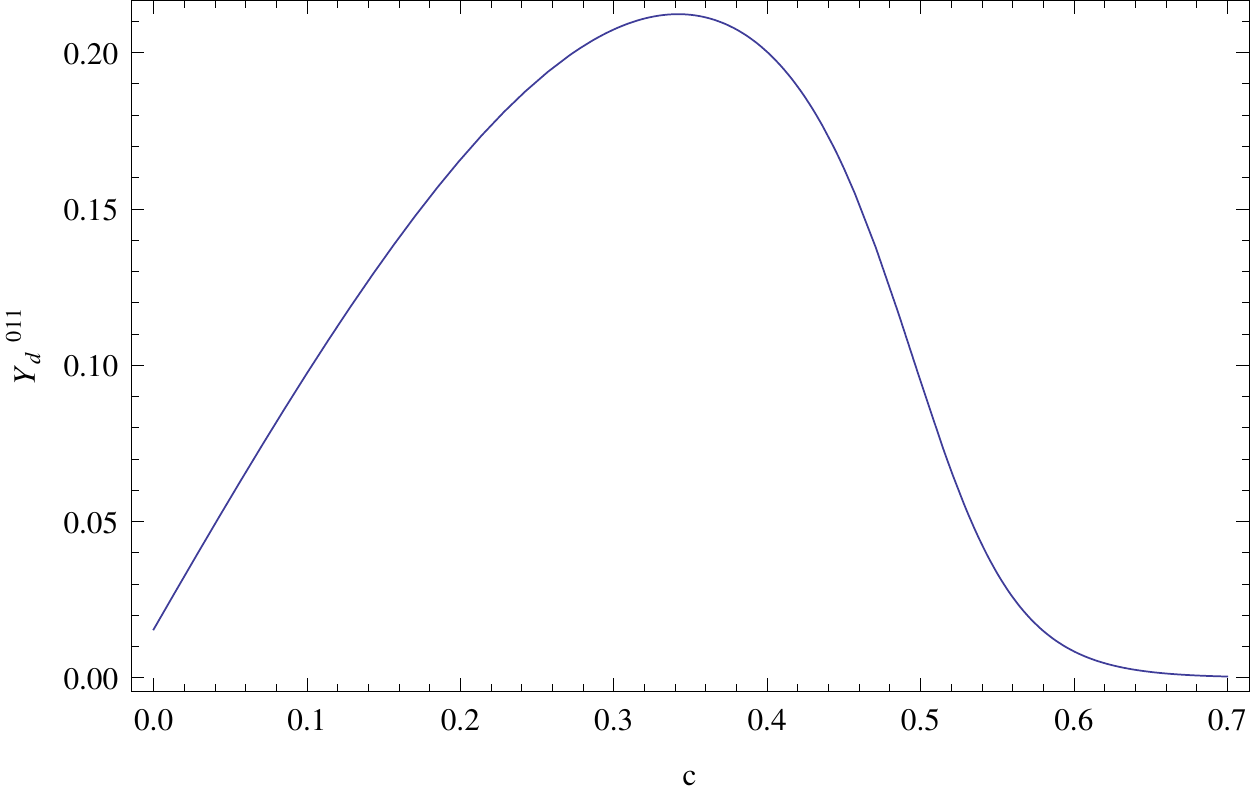}
  \end{tabular}
\caption{The figure shows the overlap ($Y^{011}_d$) between a zero mode fermion with the first KK state of $L$ and $H_d$. $c_L=0.86$ is chosen for the first KK state of $L$ }
  \label{o4}
  \end{figure}

\begin{figure}[h]
  \begin{tabular}{cc}
  \includegraphics[width=.45\textwidth,angle=0]{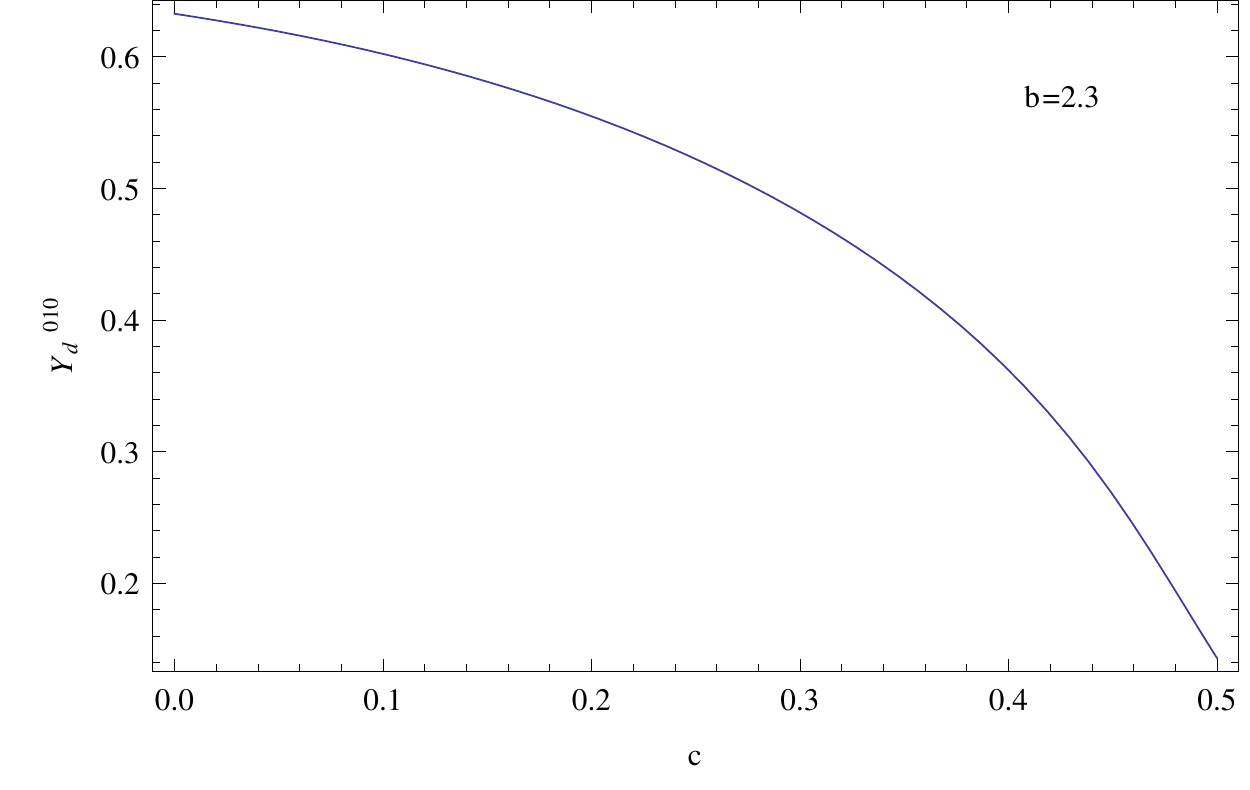}& \includegraphics[width=.5\textwidth,angle=0]{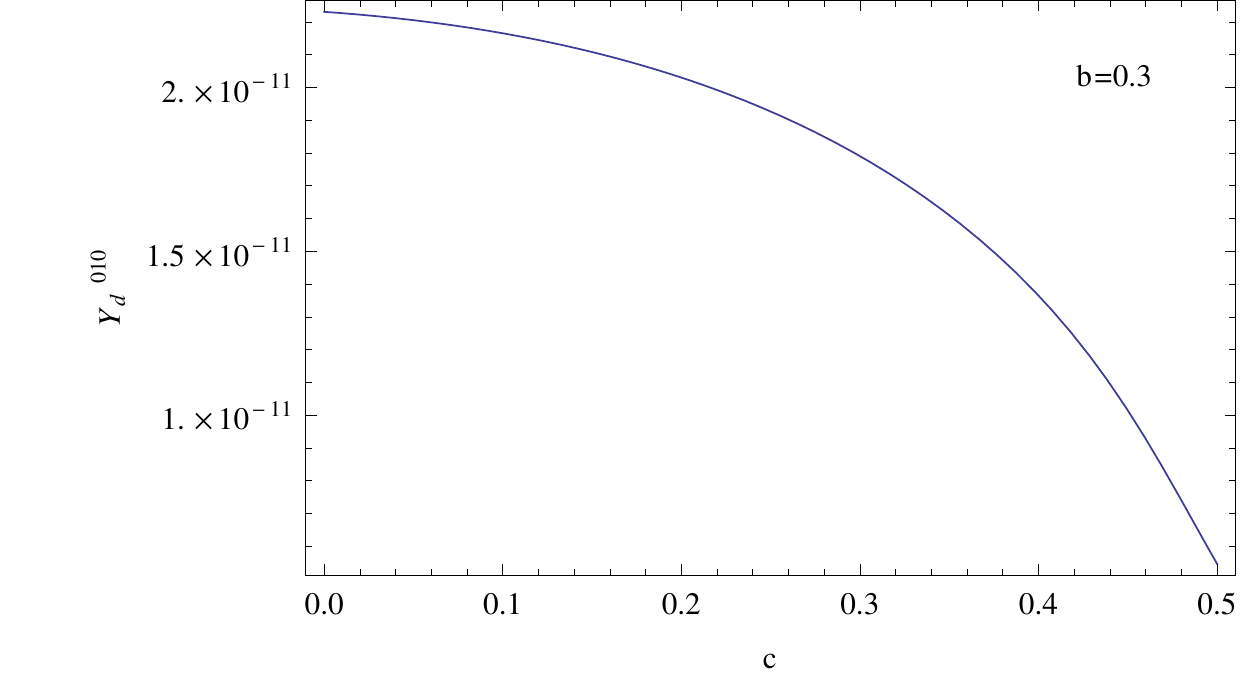}
  \end{tabular}
\caption{The figure shows the coupling $Y^{010}_d$ between a zero mode lepton, KK fermion and scalar mass eigenstate for two specific values of the brane mass parameter $b$.
The left panel corresponds to the case where the Higgs field $H_a$ is localized near the IR brane $b>1$ while the right panel corresponds to the case where it is localized
towards the UV brane $b<1$}
  \label{o3}
  \end{figure}

\begin{figure}[h]
  \begin{tabular}{c}
  \includegraphics[width=0.5\textwidth,angle=0]{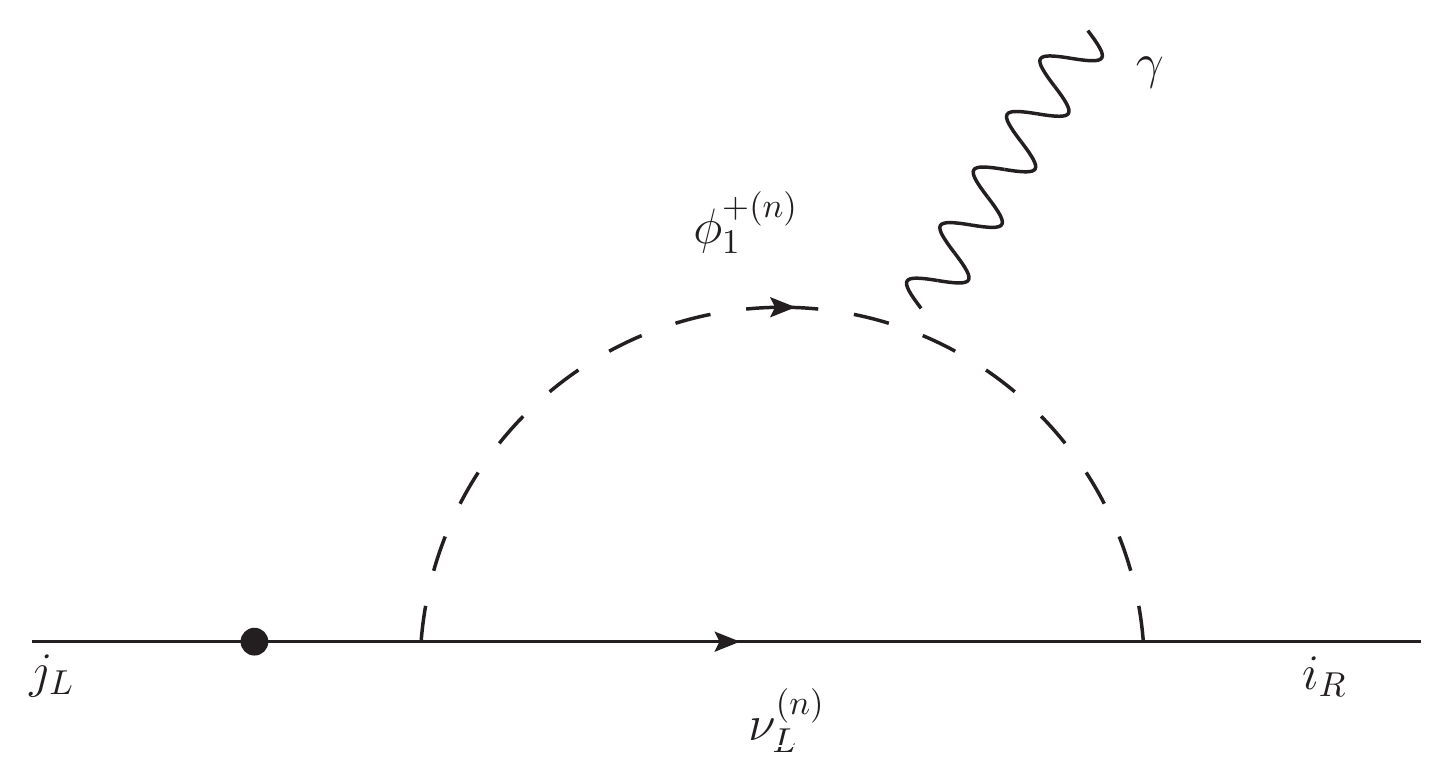}
  \end{tabular}
\caption{Figure shows the dominant contribution to  $l_j \to l_i + \gamma$ decay due to exchange of $\phi^{+(n)}$. The dot denotes mass insertion.
The localization of the first two generation charged
singlets close to the IR brane ($c<0.5$) results in $\mathcal{O}(1)$ overlap with the KK states as shown in Fig.[\ref{o4}].}
  \label{llhhdipole2}
  \end{figure}

 As shown in Fig.[\ref{o2}] the integral $g^{(n,m)}$ will be negligible for these cases since the zero mode of $H_d$ and the KK states are localized away from each other.
 In addition, consider the coupling $Y^{(0,n,0)}_{dij}$ of a zero mode fermion with flavour index i
with KK fermion state with flavour index j and a scalar mass eigenstate as defined in Appendix[\ref{couplings}]. 
The diagram on the right panel of Fig.[\ref{llhhdipole}] is proportional to $(Y^{(0,n,0)}_d)_{ij}^2$ which is negligible in this case. 
Fig.[\ref{o3}] depicts the coupling $(Y^{(0,n,0)}_d)_{ij}^2$ as a function of fermion bulk mass parameter for two different values of brane mass parameter $b_d$.
In such a scenario the main contribution arises due to the KK states of $H_d$. It will be due to the exchange of KK state of $H_d$
$\phi_1^{+(n)}$ as shown in Fig[\ref{llhhdipole2}].
There will also be a similar contribution due to $\rho_1^{(n)}$ in the loop.
Since the right handed singlets are localized near the IR brane ($c<0.5$), it's coupling to the first KK state of $H_d$ and KK fermion
will be $\mathcal{O}$(1) as shown in Fig.[\ref{o4}]. The Wilson co-efficient $A_R$ in this case is approximately given by
\begin{equation}
 A_R\sim \frac{1}{16\pi^2}\frac{(Y^{011}_d)^2m_\mu^2}{M_{KK}^2}
\end{equation}
where $Y^{011}_d$ is coupling of zero mode fermion with the first KK state of $L$ and $H_d$. Fig.[\ref{o4}] represents it's dependence on the fermionic bulk mass parameter $c$.
Since the mass insertion is on the external line, $A_L$ being suppressed by the electron mass $m_e$ is negligible compared to $A_R$. Using this expression in the branching fraction
expression in Eq.(\ref{BR}) we find that for $\mu\rightarrow e\gamma$, lowest KK scales in excess of 100 TeV is required to constrain the BR below 
the experimental upper bound of $5.7\times 10^{-13}$ \cite{Adam:2013mnn}.
 This is primarily due to the fact that the 
 coupling of the right handed singlets to the KK states of $H_d$ and $L$ is $\mathcal{O}$(1). 


  2)$H_d$ localized near the IR brane.\newline
Configurations B and D fall under this category. For configuration B the dominant contribution will be due to the exchange of $h,H,A$ in the loop. The contribution due to Fig.[\ref{llhhdipole2}]
will be smaller in comparison as the singlets and the doublets (especially for the first two generations) are predominantly localized near the UV brane ($c>0.5$) as shown in Fig:[\ref{lhlhfitb}]. 
This case also requires KK mass of several TeV to suppress
the loop induced decays( $\mu\rightarrow e\gamma$ in particular).

Configuration D, on the other hand receives contribution due to both Figures \ref{llhhdipole} as well as [\ref{llhhdipole2}]. Similar to configuration A 
very heavy KK masses $\mathcal{O}$(100) TeV is required to suppress the rates.
   
 \section{Minimal Flavour  Violation}
 \label{section6}
 As discussed in the previous section and also noted in \cite{Agashe1,perez,Iyer}, RS model with bulk fermions and gauge bosons typically give rise to large contributions to the loop induced decays
 for low lying KK scales which are within the reach of the LHC. As a result the regions which fit the lepton mass and mixing data in Fig.[\ref{llhhavailable}] are ruled
 out when constraints from the flavour sector are taken into consideration.
 One solution is to consider RS at the GUT scale where low lying KK scales are naturally of $\mathcal{O}(M_{GUT})$ thus
 giving negligible contribution to the flavour changing processes \cite{Iyer2a}. In this work we explore the alternative of invoking flavour symmetries as a solution to constrain
 the flavour changing currents. In particular we use the approach of the Minimal Flavour Violation (MFV) ansatz which was first introduced for 4D theories in the 
 hadronic sector \cite{mfv1}. According to this proposal all new sources of flavour violation are completely embedded in the Yukawa couplings of the SM. Its implementation
 in the leptonic sector is not unique owing to various modes of neutrino mass generation and various possibilities were discussed in \cite{cirigliano}.
 
 In the RS sector the large contribution to the flavour violating processes results owing to the misalignment between the mass matrix which is a function of the bulk mass parameters
 and the flavour structure of the processes which are a function of the Yukawa couplings. This can be alleviated by using the ansatz of MFV applied to the RS sector first pointed out in
 \cite{Fitzpatrick} for the quark sector. It's extensions to the leptonic was considered in \cite{perez,Chen,Iyer}. According to this ansatz the bulk Yukawa matrices can be rotated using the 
 flavour group to be aligned with the bulk mass matrices. This reduces the misalignment between the flavour basis and the mass basis thereby giving a suppressed contribution
 to the flavour changing processes.
 
 For the case under consideration the major contribution to the loop level diagrams is due to the mixing of the SM states with the KK states parametrized by $Y'_E$.
 MFV can be applied to this if we impose the following flavour symmetry on the bulk lagrangian.
 \begin{equation}
 G_{lepton}\equiv SU(3)_L\times SU(3)_{E} 
 \end{equation}
 
 The fundamental 5D Yukawa couplings have the following transformation under $G_{lepton}$:
 \begin{equation}
  Y_E \rightarrow (3,\bar 3)\;\;\;\;\;\; ; \kappa\rightarrow (6,1)
 \end{equation}
The bulk masses can be expressed in terms of the $\mathcal{O}$(1) Yukawa couplings as
 \begin{equation}
 c_L=a_1I+a_2 Y'_EY'^\dagger_E+a_3\kappa'\kappa^{'\dagger} \;\;\;\;\;c_E=bY'^\dagger_E Y'_E 
\label{align}
 \end{equation}
 Using the flavour symmetry, we can work in a basis in which $Y'_E$ is diagonal. In this basis $\kappa'$ is defined as \cite{cirigliano}
 \begin{equation}
  \kappa'\propto U^*_{PMNS}\begin{pmatrix}
                           m_{\nu_1}&0&0\\
                           0&m_{\nu_2}&0\\
                           0&0&m_{\nu_3}
                          \end{pmatrix}U^\dagger_{PMNS}
 \end{equation}
The proportionality constant is $\frac{M_{Pl}}{v^2}$. To see how the imposition of MFV affects the fermion mass fits, we provide an illustrative example for Configuration A.
We choose $c_L=0.87$ for all three generations. To fit the charged lepton masses we choose $c_E=\text{Diag}(0.256,0.418,0.56)$. The corresponding $\mathcal{O}$(1) Yukawa
couplings are $Y'_E=\text{Diag}(0.69,0.875,1.02)$. In this case the parameter $b$ in Eq.(\ref{align}) is chosen to be 0.538. 

Similarly we provide an illustrative example for Configuration D. We choose $c_L=0.87$ for all three generations. The charged lepton masses by choosing $c_E=\text{Diag}(0.2407,0.427,0.85)$. The corresponding $\mathcal{O}$(1) Yukawa
couplings are $Y'_E=\text{Diag}(1.26,1.66,2.29)$. The parameter $b$ in this is chosen to be $\sim 0.153$. Similar examples can be obtained for the other two cases as well. 
On comparing with the leptonic fits in Table[\ref{ranges}], we find that the imposition of MFV does not change the fundamental nature of fits. The only requirement is that the 
$c$ parameters chosen for the fit must be proportional to the Yukawa couplings.

We can construct a flavour violating combination ($\Delta$) transforming as (8,1) under $G_{lepton}$ as
\begin{equation}
 \Delta=\kappa^{'\dagger}\kappa'
\end{equation}
The higher dimensional operator invariant under $G_{lepton}$ which parametrizes $j\rightarrow i\gamma$ is given as
\begin{equation}
 \mathcal{O}_{j\rightarrow i\gamma}=e_5H_d^\dagger Y_E\bar E_{i}\sigma_{\mu\nu}\Delta L_jF^{\mu\nu}
\end{equation}
where the fields are bulk fields.
The contribution of flavour changing diagrams (parameterized by $\Delta$) will be proportional to $\frac{v^2}{M^2_{PL}}$ as a result of the higher dimensional operator being suppressed
by the Planck scale. Thus as a result of implementing the MFV ansatz the dangerous flavour violating contributions are highly suppressed. 
 \section{Conclusions} 
 \label{section7}
 Neutrino masses through quantum gravity effects is an interesting possibility. In an effective theory this is represented by the LLHH operator.
Phenomenologically the LHLH scenario with large warp factor and a single Higgs doublet is known to give an unsatisfactory fit to the data owing to the large negative values of the
$c_E$ parameters. Such scenarios can however be effectively rescued by considering Higgs fields in the bulk. In particular, models with two Higgs doublets of Type II
were considered which are useful when supersymmetric models are considered.
Four configuration of Higgs fields in the bulk were considered.
The neutrino and charged lepton masses as well as the neutrino mixing angles for all the four configurations, can
be fit with $\mathcal{O}$(1) choice of bulk parameters. Constraints from flavour are very severe often requiring the the lowest KK scales to be very heavy.
Minimal Flavour Violation (MFV) is a useful tool which helps in controlling the branching fractions without the introduction of heavy scales in the theory.
We provide an example for the implementation of the MFV scenario for the case under consideration.
The scenario of RS with two Higgs doublets is still in its infancy. It could be interesting to consider various issues
relating to vacuum stability, electroweak precision constraints \textit{etc.} which could establish such models on a much firmer footing.

\vskip 1 cm
 \noindent
\textbf{Acknowledgement}\\

I am grateful to Sudhir Vempati for his encouragement, numerous discussions and his suggestions regarding the manuscript.
I would like to thank Daisuke Harada for clarifying a point regarding
two Higgs doublet models. 
In addition I would also like to acknowledge useful discussions with V.Surayanarayana Mummidi, Debtosh Chowdhury and Kirtimaan Mohan on EWSB in supersymmetric scenarios. 
I am thankful to the Indian tax payer and the Govt. of India for promoting research in basic sciences. \newline

\newpage
\appendix
\section{Parameter space of bulk masses}
\label{plots}
 \begin{figure}[htp]
\begin{tabular}{cc}
\includegraphics[width=0.4\textwidth,angle=0]{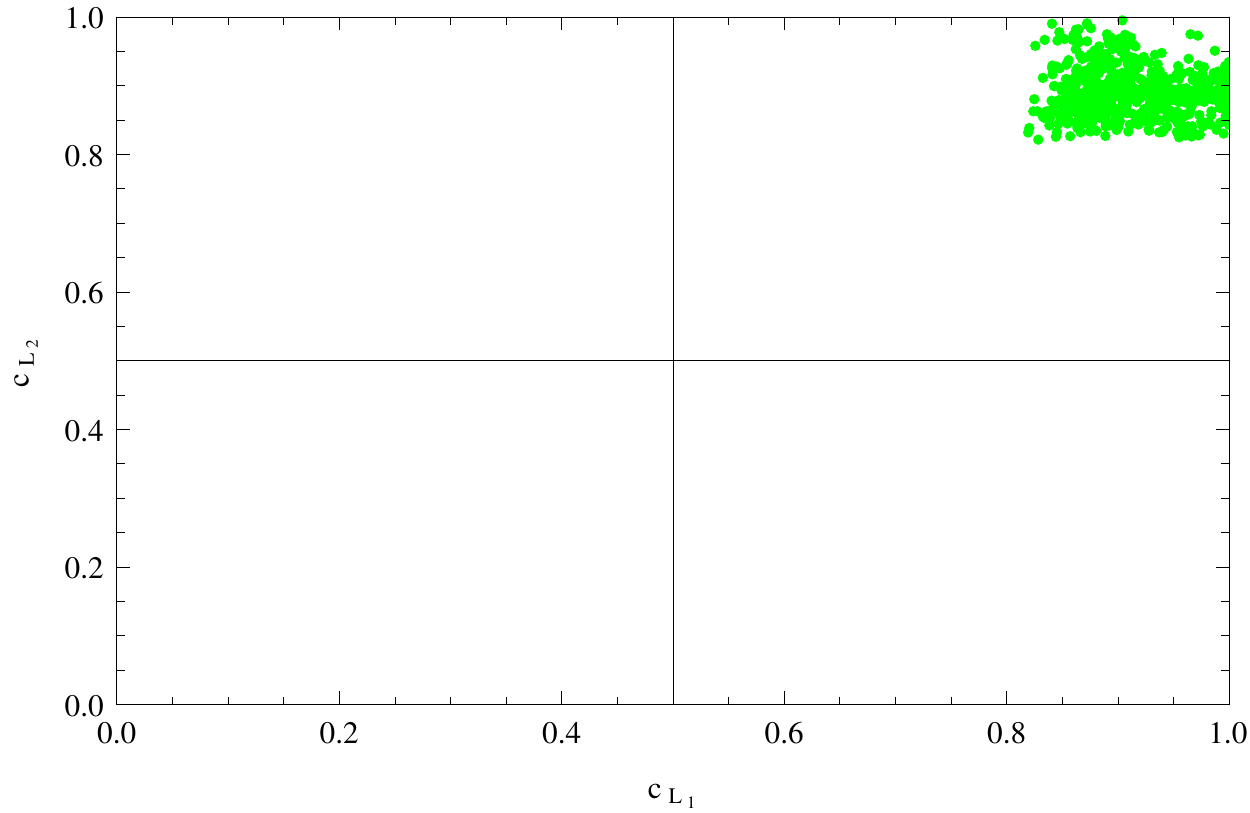} &
\includegraphics[width=0.4\textwidth,angle=0]{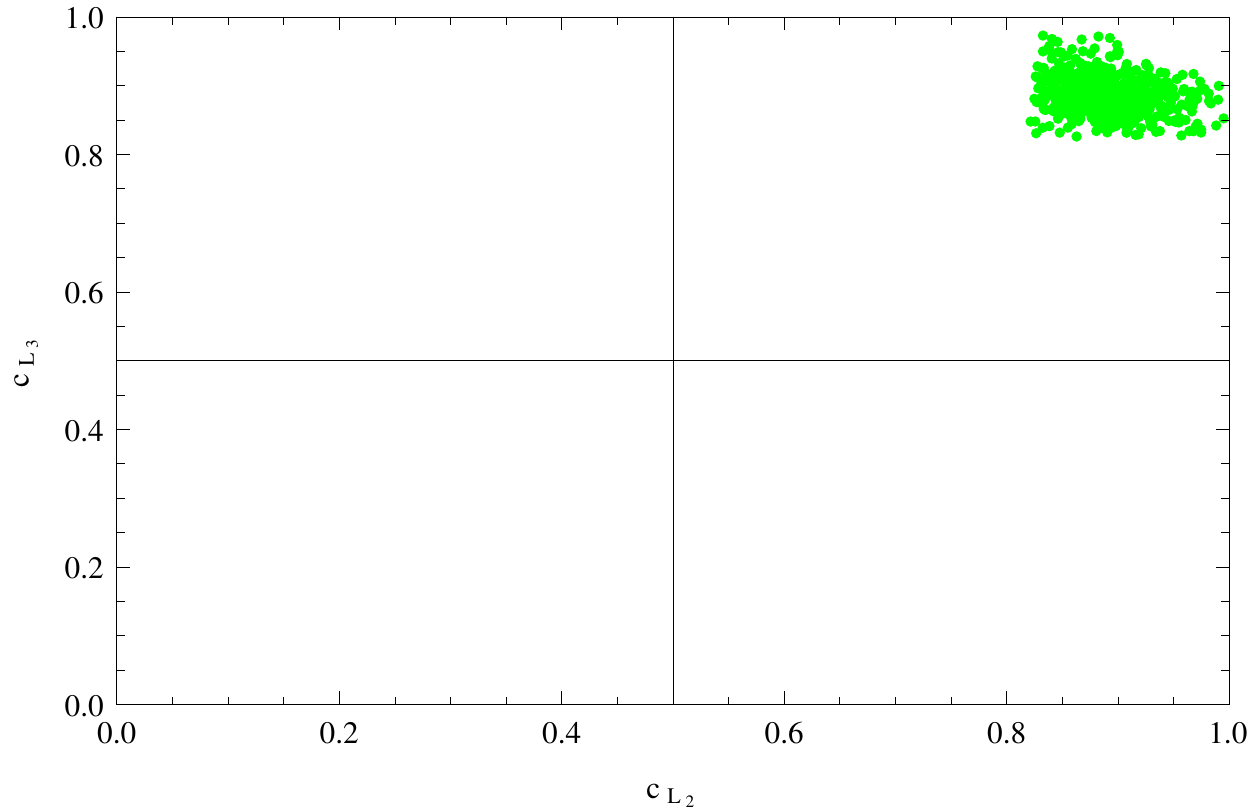} \\
\includegraphics[width=0.4\textwidth,angle=0]{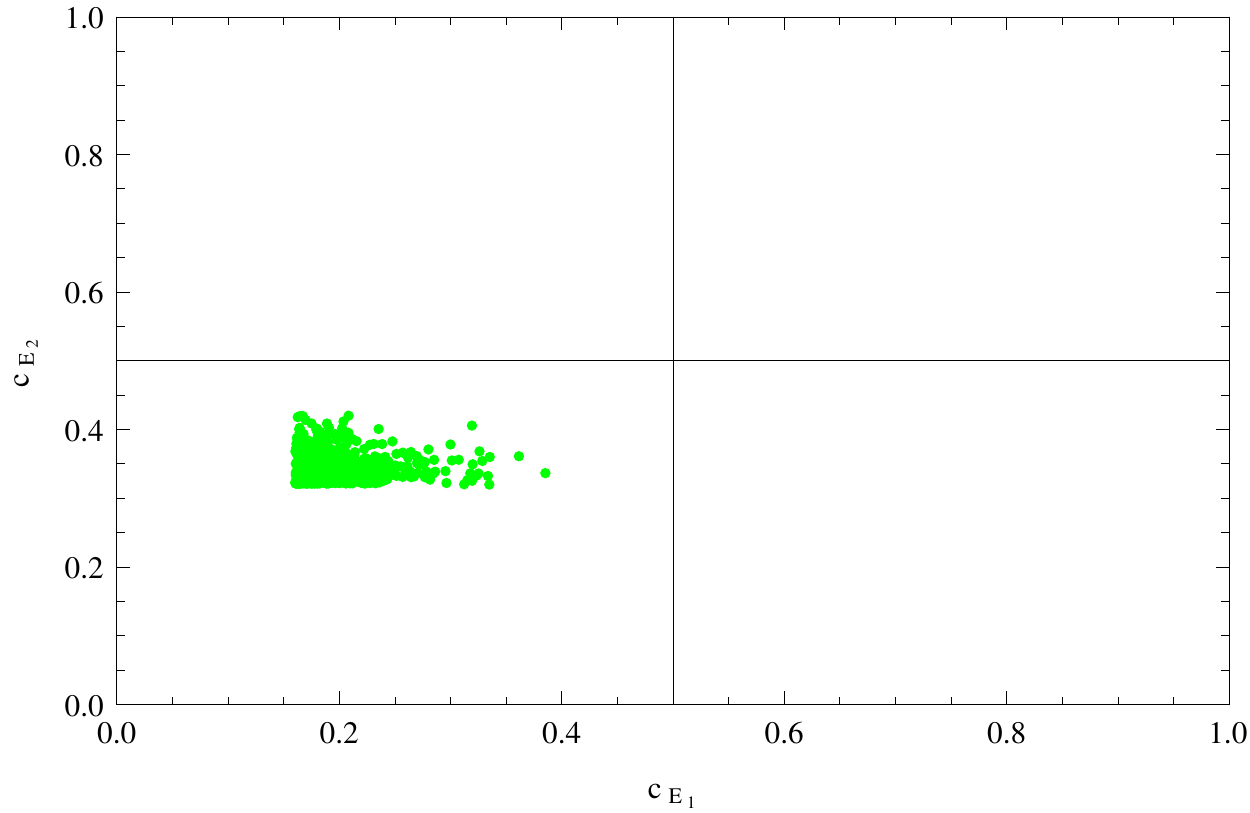} &
\includegraphics[width=0.4\textwidth,angle=0]{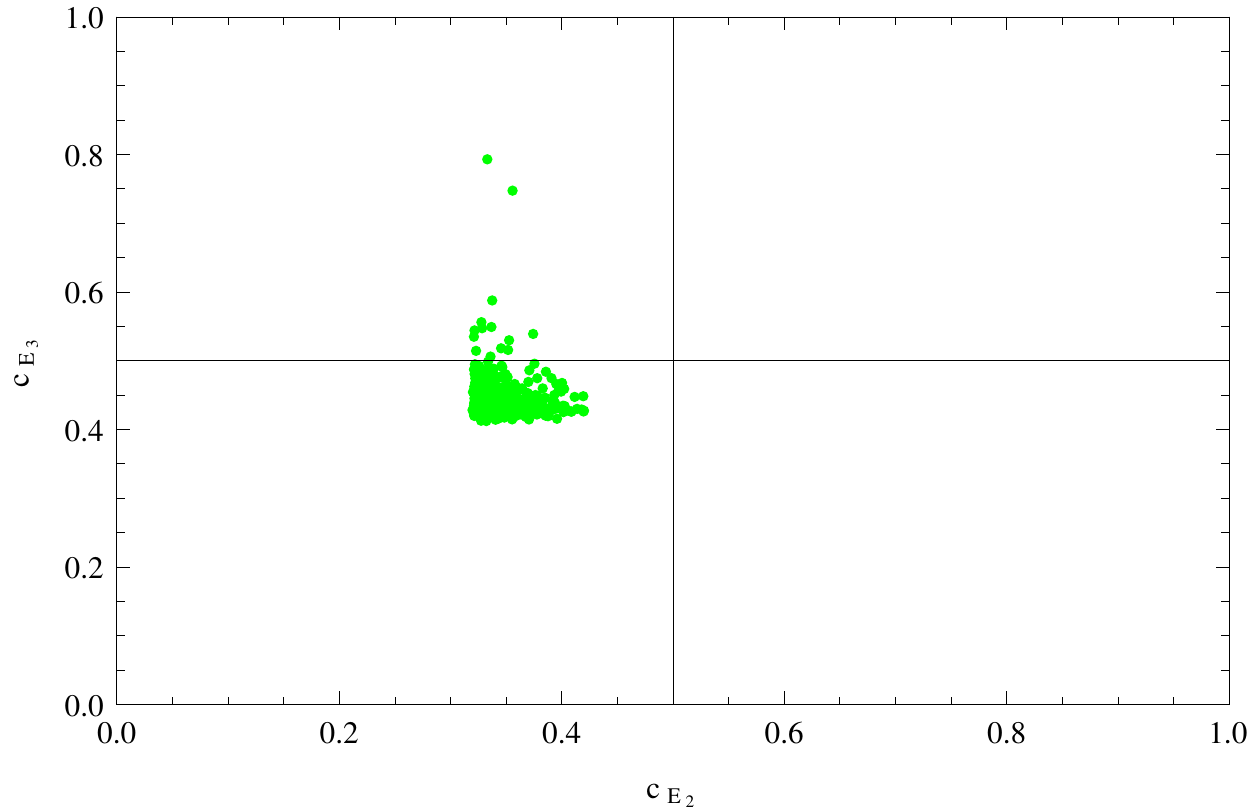}
\end{tabular}
 \caption{Parameter space of bulk masses for configuration A. The graphs in the upper row shows the parameter space for
 the bulk masses for doublets ($c_{L_{i}}$) while the lower row shows the corresponding parameter space for for the charged singlets  $c_{E_i}$. 
 All points satisfy $0<\chi^2<10$.}
\label{lhlhfita}
\end{figure}
\begin{figure}[htp]
\begin{tabular}{cc}
\includegraphics[width=0.4\textwidth,angle=0]{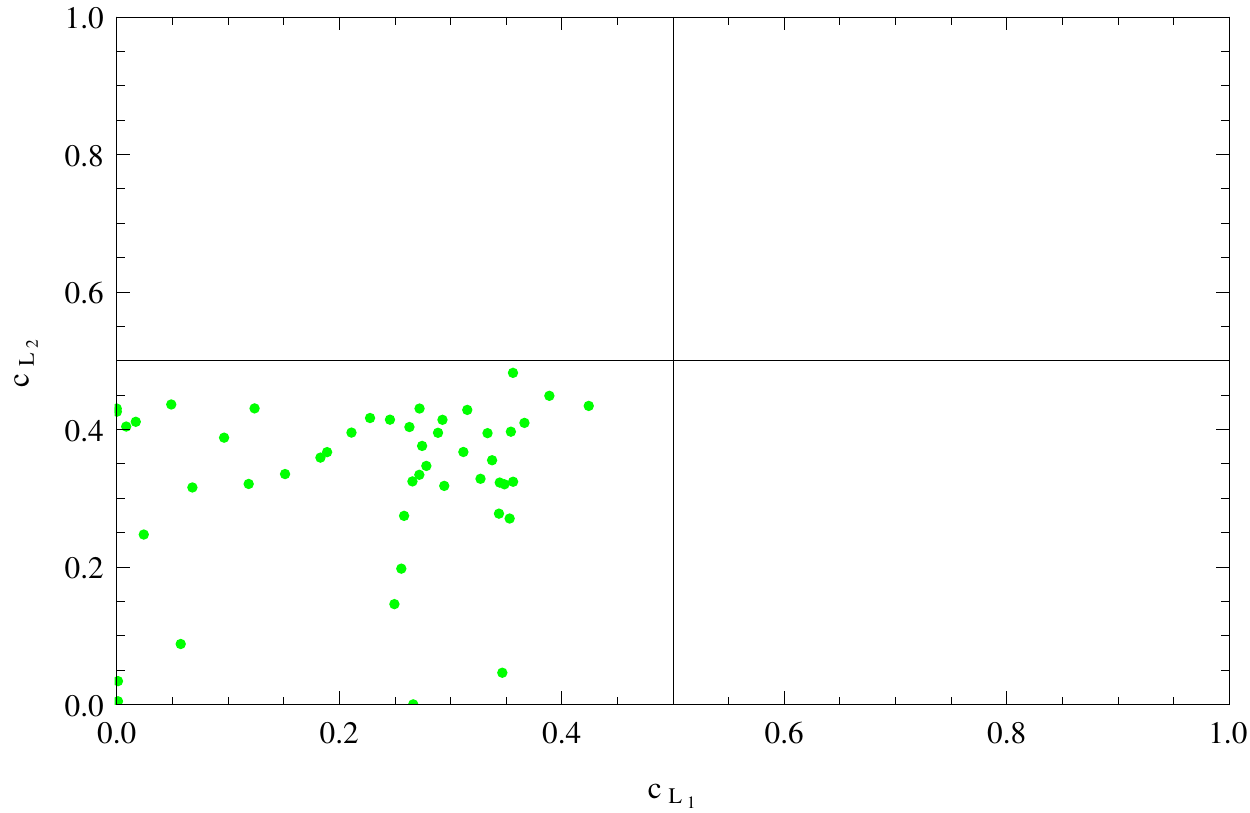} &
\includegraphics[width=0.4\textwidth,angle=0]{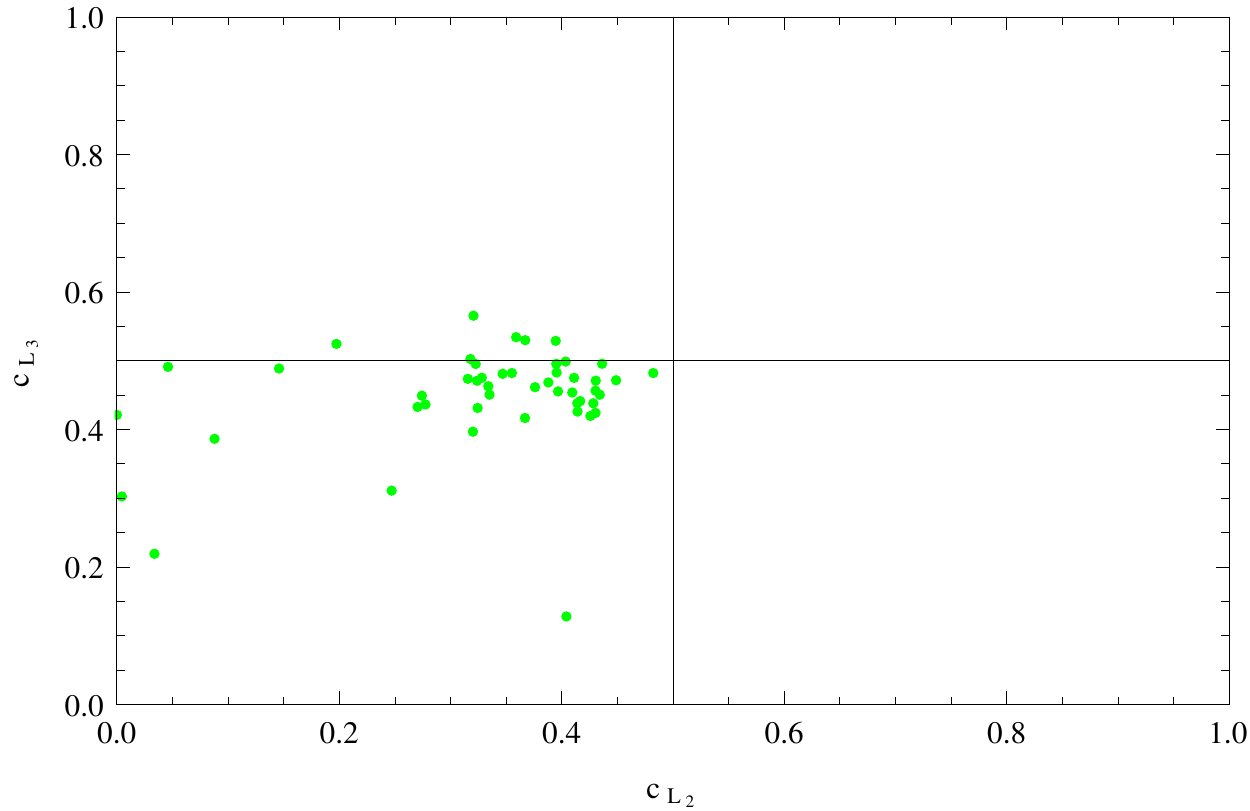} \\
\includegraphics[width=0.4\textwidth,angle=0]{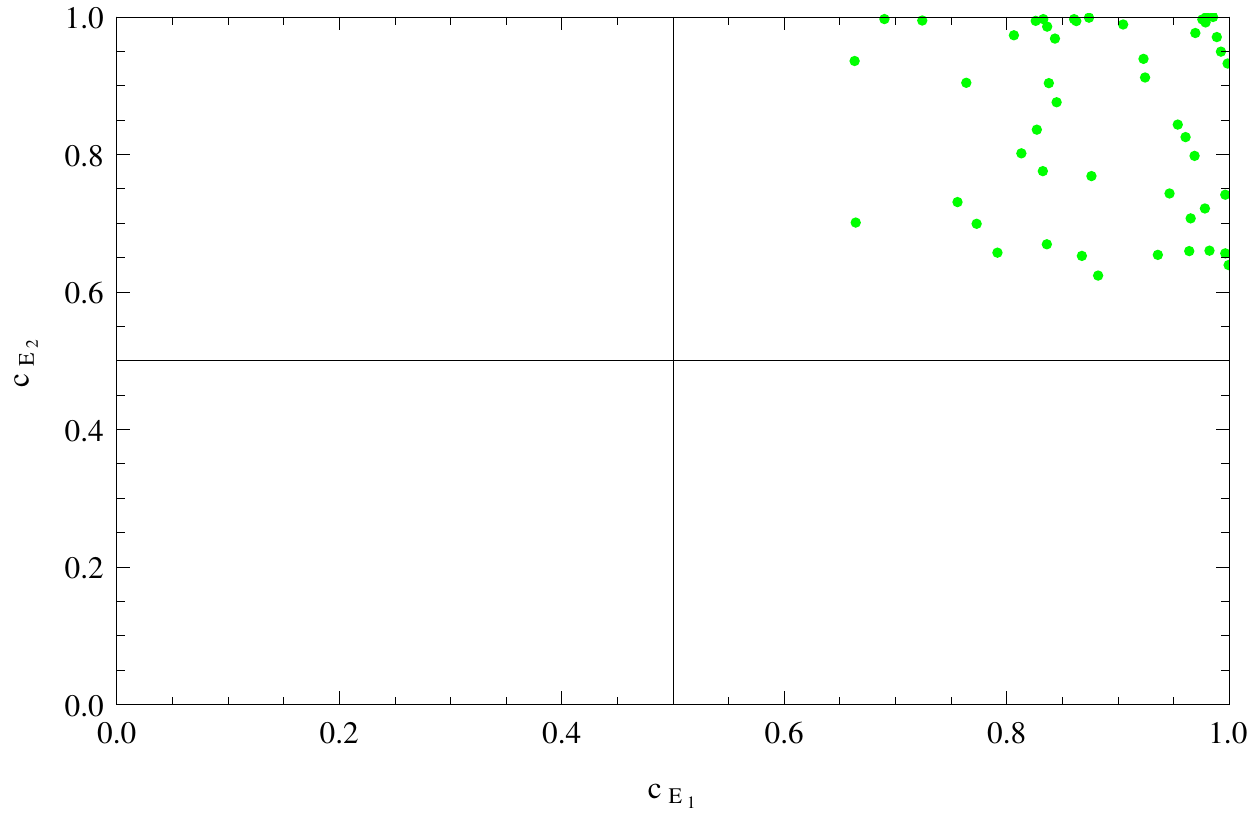} &
\includegraphics[width=0.4\textwidth,angle=0]{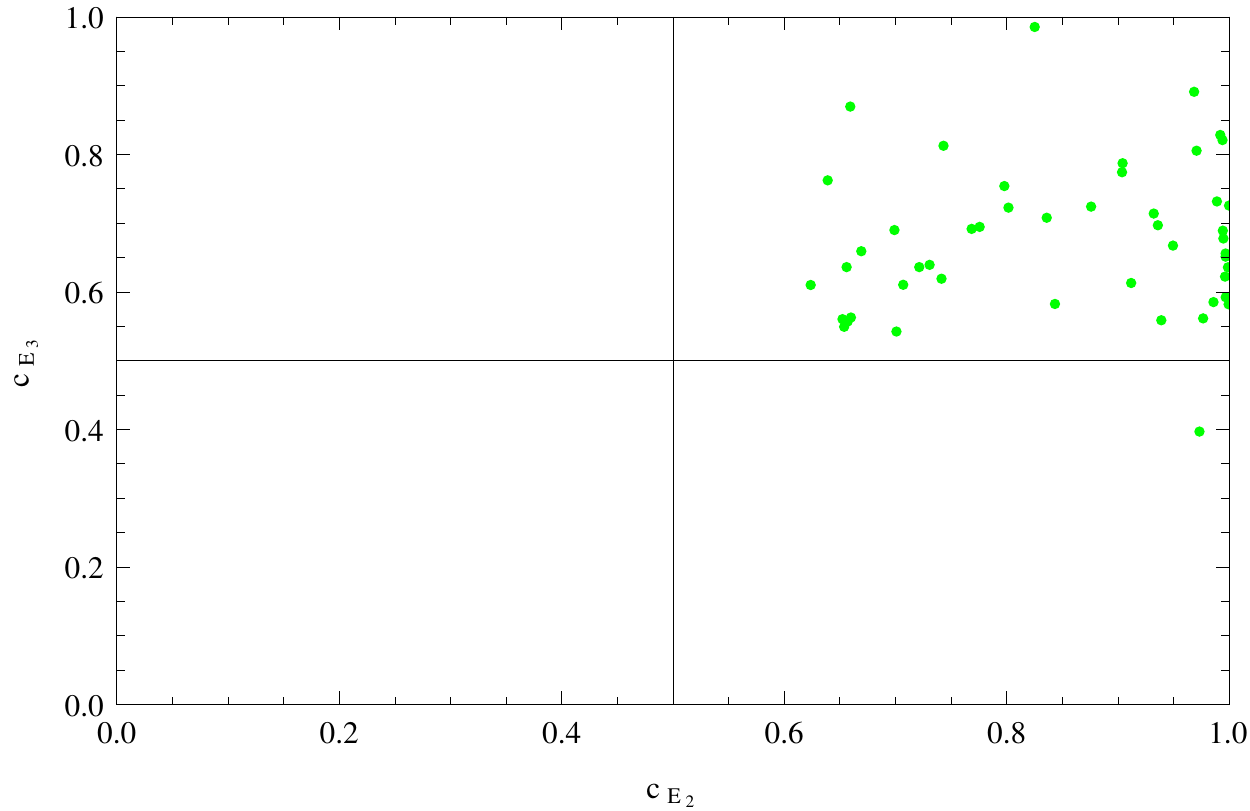}\\
\end{tabular}
 \caption{Parameter space of bulk masses for configuration B.}
\label{lhlhfitb}
\end{figure}
 \newpage
\begin{figure}[htp]
\begin{tabular}{cc}
\includegraphics[width=0.4\textwidth,angle=0]{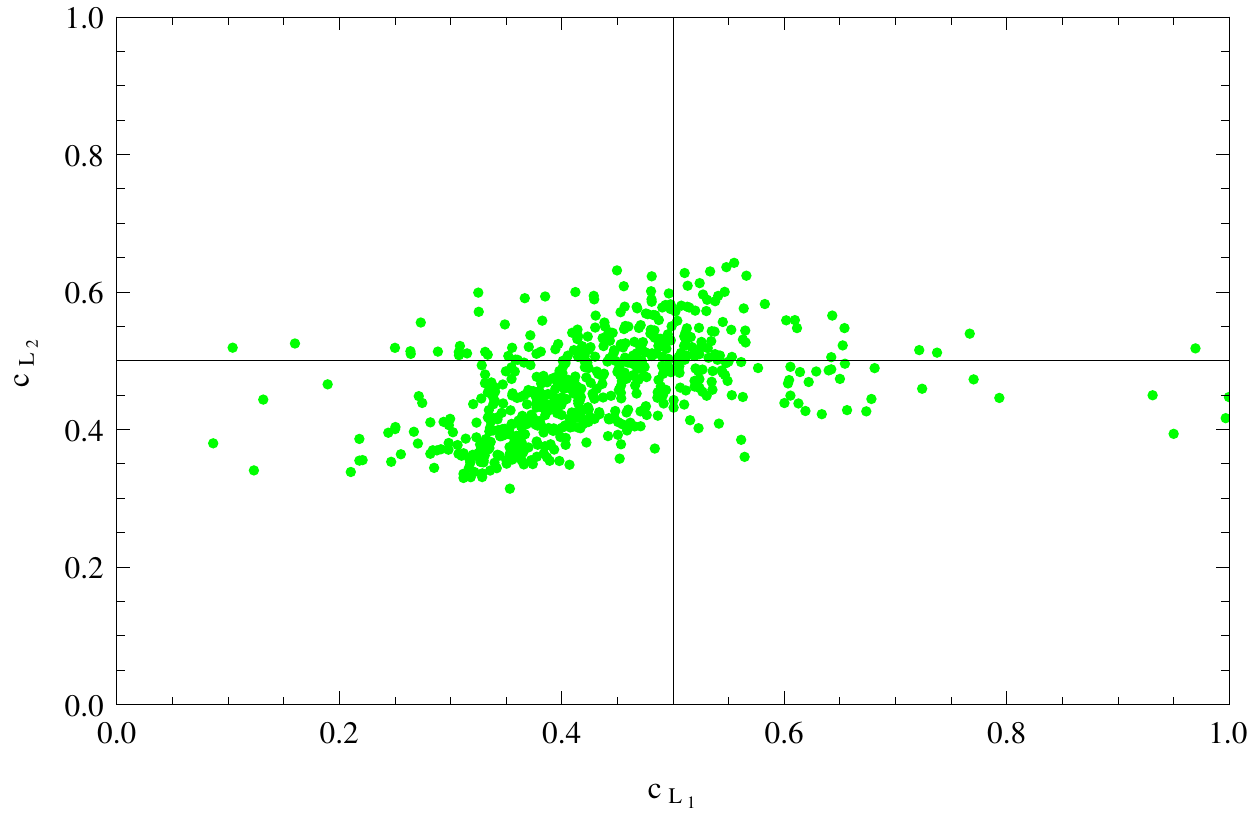} &
\includegraphics[width=0.4\textwidth,angle=0]{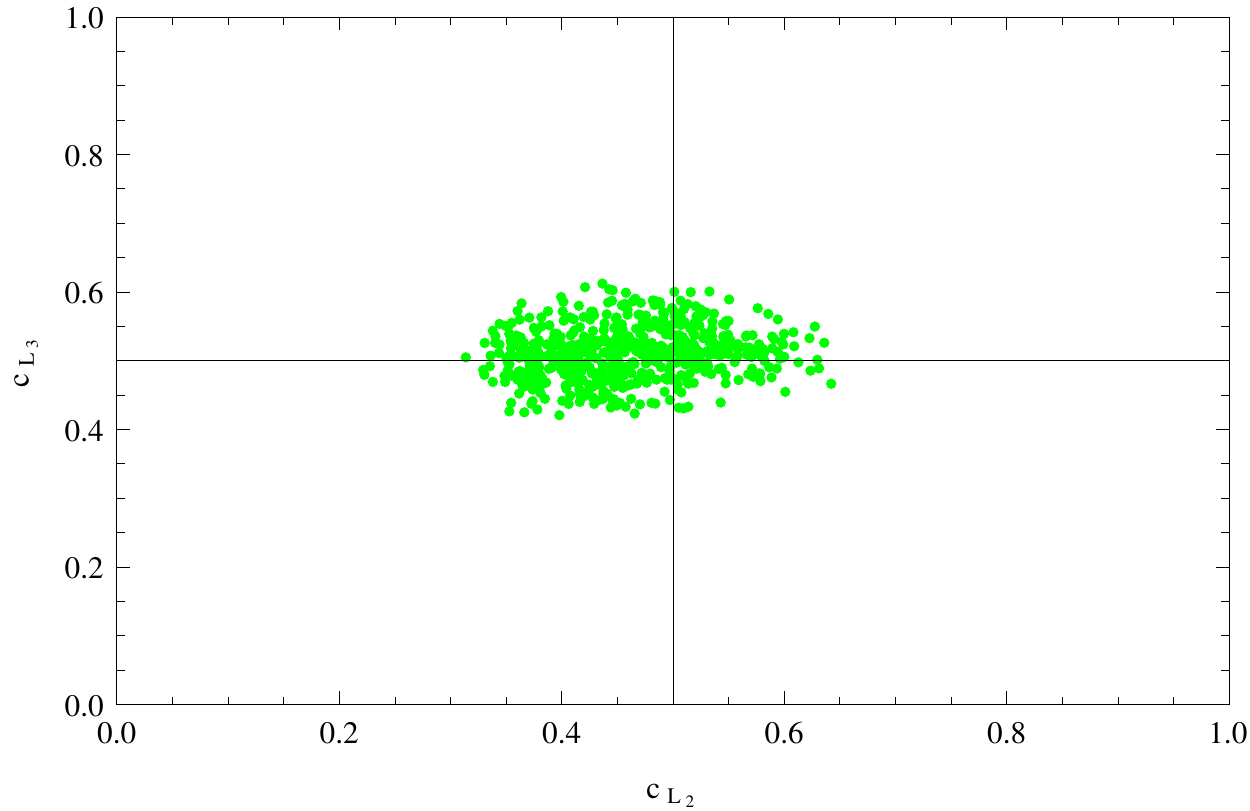} \\
\includegraphics[width=0.4\textwidth,angle=0]{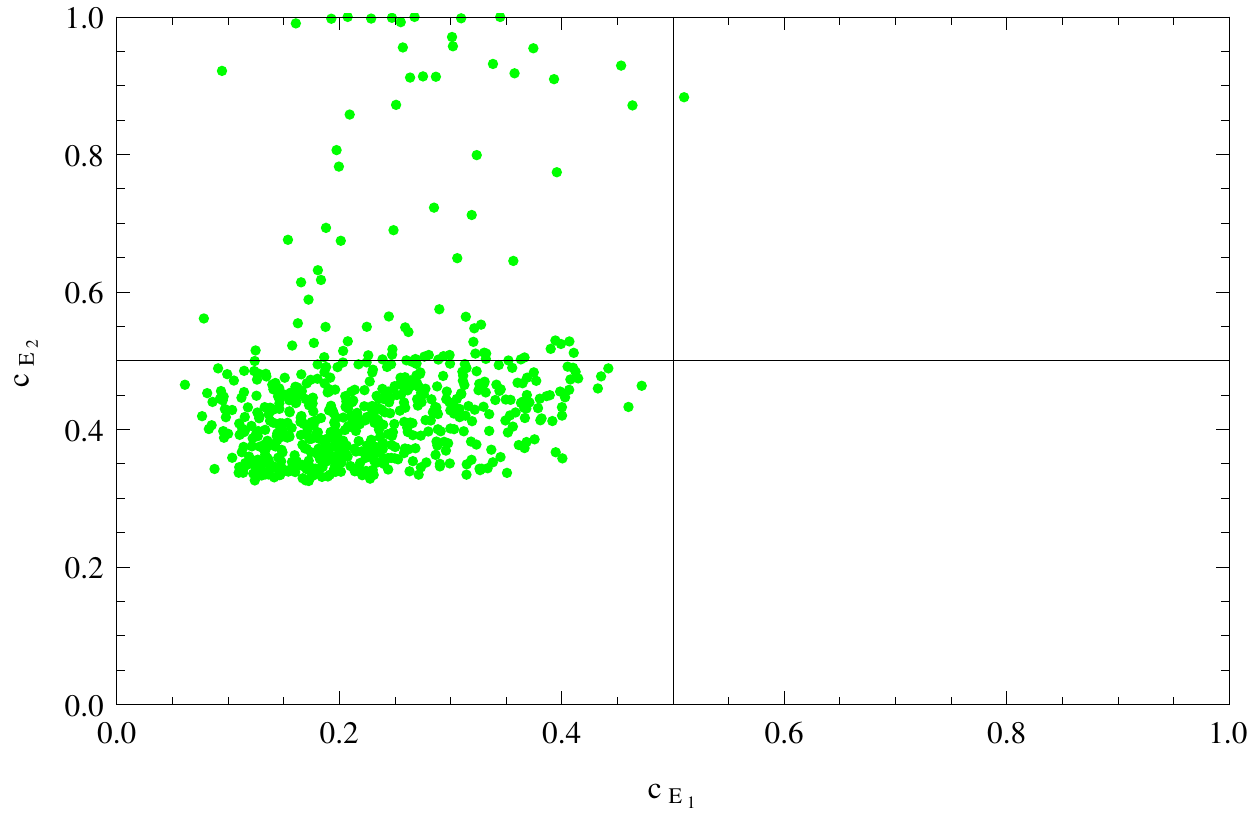} &
\includegraphics[width=0.4\textwidth,angle=0]{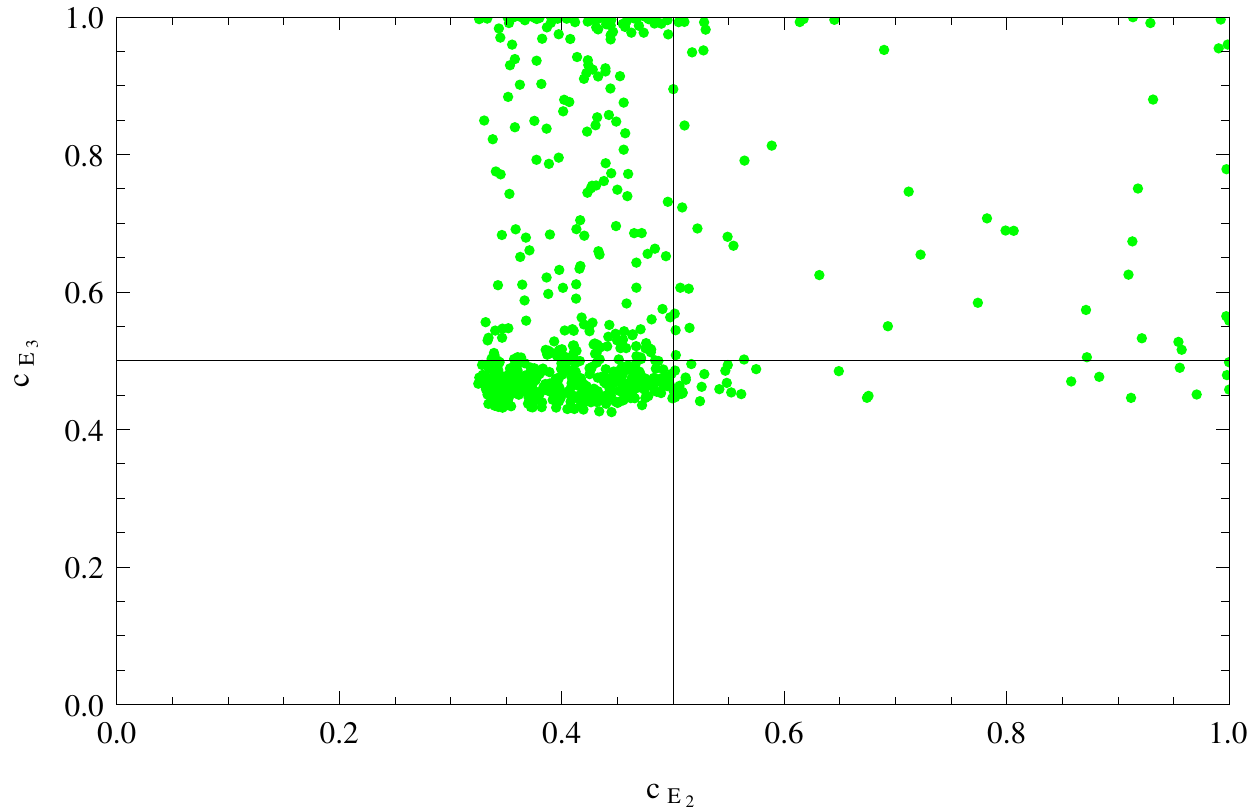}
\end{tabular}
 \caption{Parameter space of bulk masses for configuration C.}
\label{lhlhfitc}
\end{figure}
\begin{figure}[htp]
\begin{tabular}{cc}
\includegraphics[width=0.4\textwidth,angle=0]{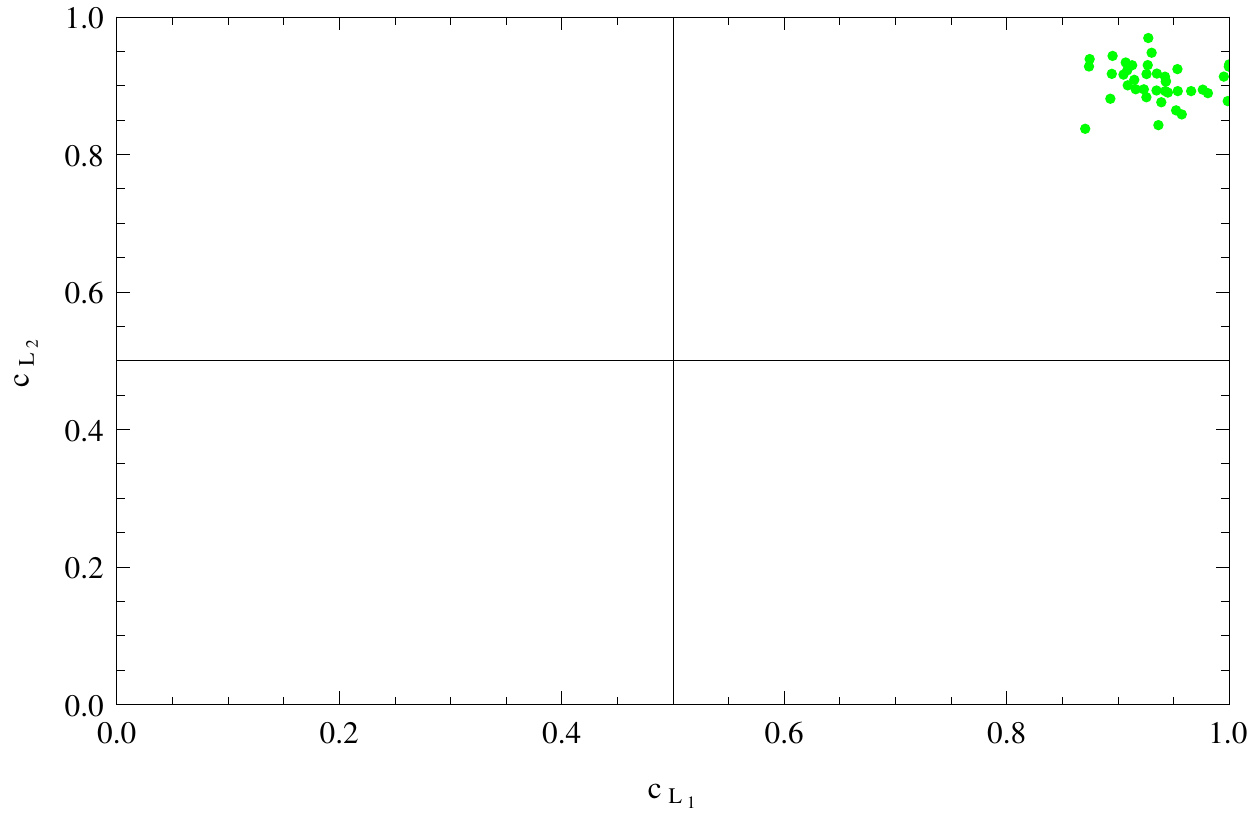} &
\includegraphics[width=0.4\textwidth,angle=0]{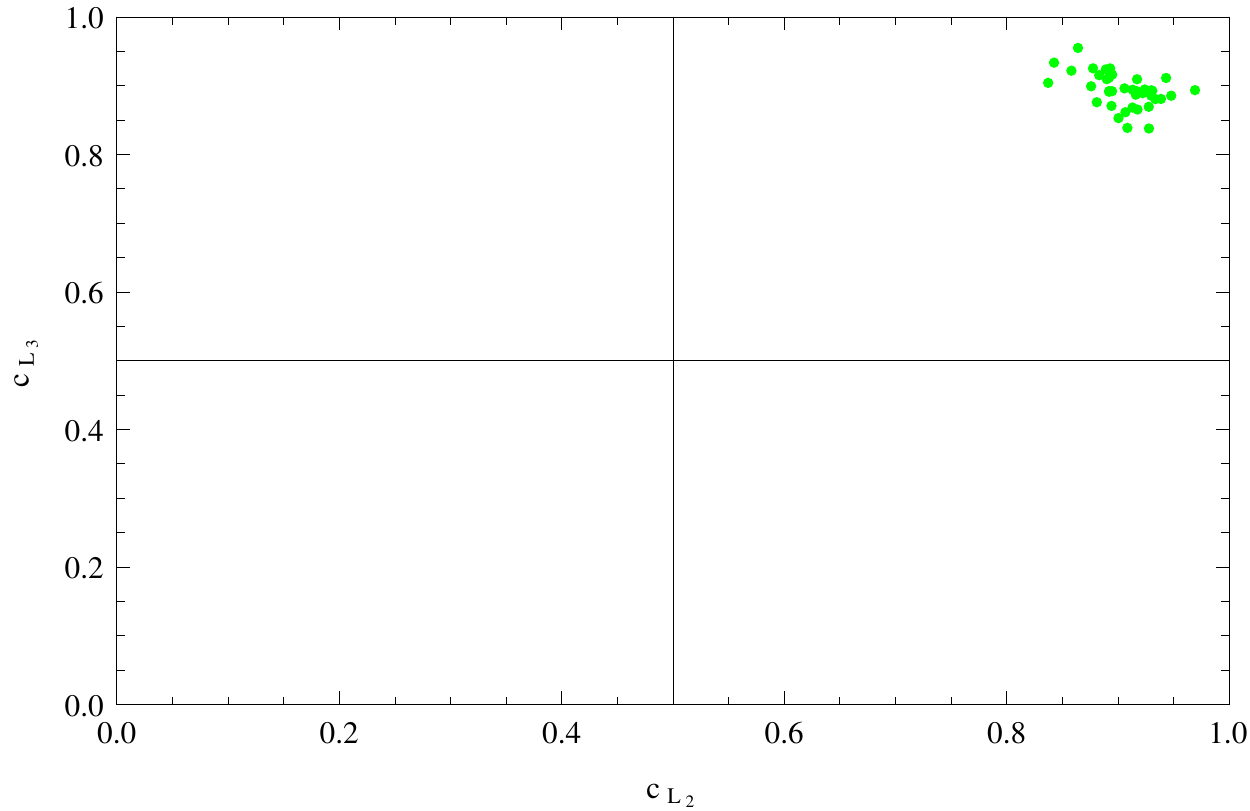} \\
\includegraphics[width=0.4\textwidth,angle=0]{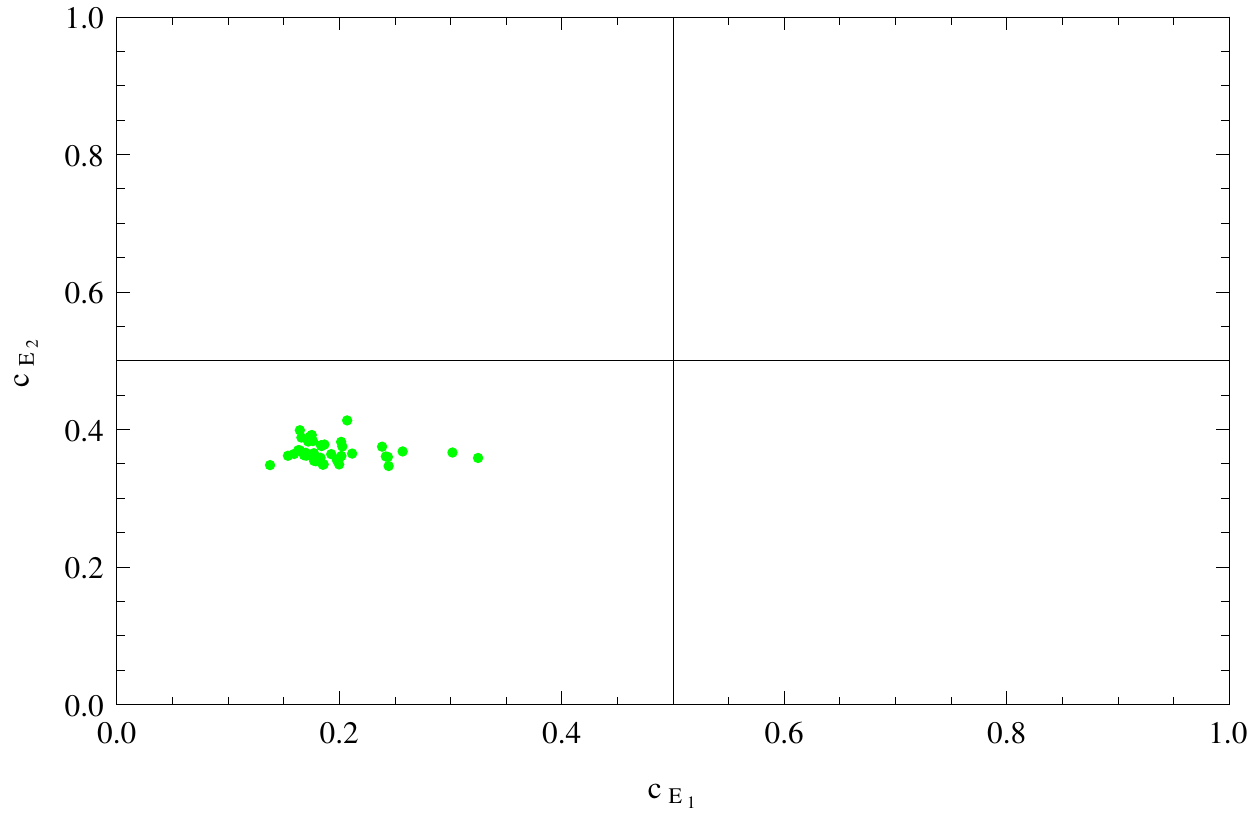} &
\includegraphics[width=0.4\textwidth,angle=0]{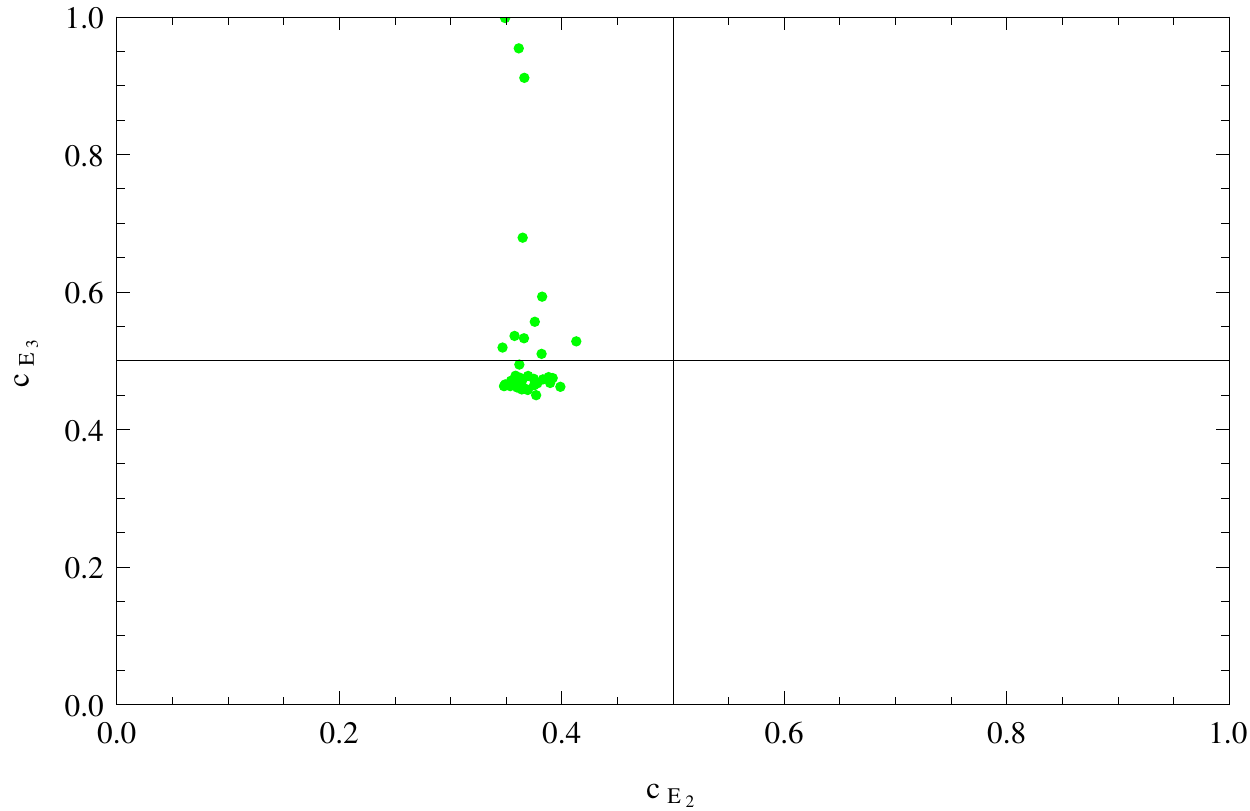}
\end{tabular}
 \caption{Parameter space of bulk masses for configuration D.
}
\label{lhlhfitd}
\end{figure}
\section{Two Higgs doublet potential in RS}
\label{couplings}
The zero mode doublets can be expressed as:
\begin{equation}
 H^{(0)}_u=\begin{pmatrix}
        \phi^{+(0)}_2\\\frac{v_2+\rho^{(0)}_2+i\eta^{(0)}_2}{\sqrt{2}}     
     \end{pmatrix}\;\;\;\;\;H^{(0)}_d=\begin{pmatrix}
        \phi^{+(0)}_1\\\frac{v_1+\rho^{(0)}_1+i\eta^{(0)}_1}{\sqrt{2}}     
     \end{pmatrix}
\label{zero modedoublets}
     \end{equation}
Three of these fields are unphysical as they are eaten up by the W and Z gauge bosons resulting in 5 physical mass eigenstates: 2 CP even, 1 CP odd and 2 charged Higgs.
The mass eigenstates are related to the interaction eigenstates by the following transformation:
\begin{equation}
 \begin{pmatrix}
  \phi_1^+\\\phi_2^+
 \end{pmatrix}=O(\beta)\begin{pmatrix}
  G^+\\H^+
 \end{pmatrix}\;\;\;\;,\begin{pmatrix}
  \eta_1\\\eta_2
 \end{pmatrix}=O(\beta)\begin{pmatrix}
  G^{(0)}\\A
 \end{pmatrix}\;\;\;\;,\begin{pmatrix}
  \rho_1\\\rho_2^+
 \end{pmatrix}=O(\alpha)\begin{pmatrix}
  h\\H
 \end{pmatrix}
\end{equation}
where 
\begin{equation}
 O(\theta)=\begin{pmatrix}
  \cos\theta&-\sin\theta\\\sin\theta&\cos\theta
 \end{pmatrix}
\end{equation}
The coupling of the SM fermions to the Higgs mass eigenstates is exactly the same for a general type II two Higgs doublet model in 4D and is given in\cite{Aoki,Branco}.
The coupling of a zero mode and KK mode charged fermion with the CP even scalar eigenstates is given as
\begin{eqnarray}
 \mathcal{L}&=&D_{Lik}\frac{Y_{dkj}^{(0,1,0)}}{\sqrt{2}}\bar d_{L_i}^{(0)}\left[Hc_\alpha-hs_\alpha\right]d_{R_j}^{(1)}+\frac{Y_{dik}^{(1,0,0)}}{\sqrt{2}}D_{Rkj}^\dagger\bar d_{L_i}^{(1)}\left[Hc_\alpha-hs_\alpha\right]d_{R_j}^{(0)}\nonumber\\
            &+&U_{Lik}\frac{Y_{ukj}^{(0,1,0)}}{\sqrt{2}}\bar u_{L_i}^{(0)}\left[Hs_\alpha+hc_\alpha\right]u_{R_j}^{(1)}+\frac{Y_{uik}^{(1,0,0)}}{\sqrt{2}}U_{Rkj}^\dagger\bar u_{L_i}^{(1)}\left[Hs_\alpha+hc_\alpha\right]u_{R_j}^{(0)} 
\end{eqnarray}
where $U_{L,R}(D_{L,R})$ are the left and right rotation matrices for the up(down) sector. The overlap matrices $Y^{(n,m,k)}_{aij}$, which parametrizes the coupling
of $n^{th}$ and $m^{th}$ KK mode of left and right  chiral fermion respectively with the $k^{th}$ KK mode of $H_d$ 
are defined as
\begin{equation}
 Y^{(n,m,k)}_{a(ij)}=\frac{Y'}{(k\pi R)^{3/2}}\int dy~\sqrt{-g}f^{(n)}_{L_i}(y)f^{(m)}_{R_j}(y)f^{(k)}_{H_a}(y)\;\;\;\;\;\;\ \text{$a=u,d$}
\label{nmk}
 \end{equation}
Here $Y'=2\sqrt{k}$ in general denotes the $\mathcal{O}$(1) of the fermions to the Higgs.
Similarly the coupling to the CP odd eigenstate is given as
\begin{eqnarray}
 \mathcal{L}=&=&-D_{Lik}\frac{Y_{dkj}^{(0,1,0)}}{\sqrt{2}}\bar d_{L_i}^{(0)}As_\beta d_{R_j}^{(1)}+\frac{Y_{dik}^{(1,0,0)}}{\sqrt{2}}D_{Rkj}^\dagger\bar d_{L_i}^{(1)}As_\beta d_{R_j}^{(0)}\nonumber\\
            &-&U_{Lik}\frac{Y_{ukj}^{(0,1,0)}}{\sqrt{2}}\bar u_{L_i}^{(0)}Ac_\beta u_{R_j}^{(1)}+\frac{Y_{uik}^{(1,0,0)}}{\sqrt{2}}U_{Rkj}^\dagger\bar u_{L_i}^{(1)}A c_\beta u_{R_j}^{(0)} 
\end{eqnarray}
 and the charged Higgs is given as
 
 \begin{eqnarray}
 \mathcal{L}=&=&-D_{Lik}\frac{Y_{dkj}^{(0,1,0)}}{\sqrt{2}}\bar d_{L_i}^{(0)}H^-s_\beta u_{R_j}^{(1)}-\frac{Y_{dik}^{(1,0,0)}}{\sqrt{2}}U_{Rkj}^\dagger\bar d_{L_i}^{(1)}H^-s_\beta u_{R_j}^{(0)}\nonumber\\
            &-&U_{Lik}\frac{Y_{ukj}^{(0,1,0)}}{\sqrt{2}}\bar u_{L_i}^{(0)}H^+c_\beta d_{R_j}^{(1)}+\frac{Y_{uik}^{(1,0,0)}}{\sqrt{2}}D_{Rkj}^\dagger\bar u_{L_i}^{(1)}H^+ c_\beta d_{R_j}^{(0)} 
\end{eqnarray}
\section{Mixing of zero mode and KK modes of $H_a$ ($a=u,d$)}
\label{mixingbetween}
The presence of interaction terms in the bulk, leads to the mixing of the zero mode  and higher KK modes of the bulk scalar field $H_a$. For simplicity will will consider only one
KK mode. We will consider the case of $H_d$ and exactly similar argument will follow for $H_u$ as well.
The $2\times 2$ mass matrix which illustrates this mixing for the charged component of $H_d$, in the basis $\left(\phi^{+(0)}_1,\phi^{+(1)}_1\right)^T$ is given as
\begin{equation}
\mathcal{M}_{\phi^{+(0)}_1,\phi^{+(1)}_1}= \begin{pmatrix}
  m_{12}^2\frac{v_2}{v_1}K^{(0,0)}_{ud}-\frac{\lambda'_4+\lambda'_5}{2}v_2^2G^{(0,0)}_{ud}&\frac{\lambda'_1}{2}v_1^2H^{0001}_{1111}+\frac{\lambda'_3}{2}v_2^2H^{0001}_{1122}\\
  \frac{\lambda'_1}{2}v_1^2H^{0001}_{1111}+\frac{\lambda'_3}{2}v_2^2H^{0001}_{1122}&M_{KK}^2+\mathcal{0}(v^2)
 \end{pmatrix}
\end{equation}
\begin{figure}[h]
  \begin{tabular}{c}
  \includegraphics[width=0.5\textwidth,angle=0]{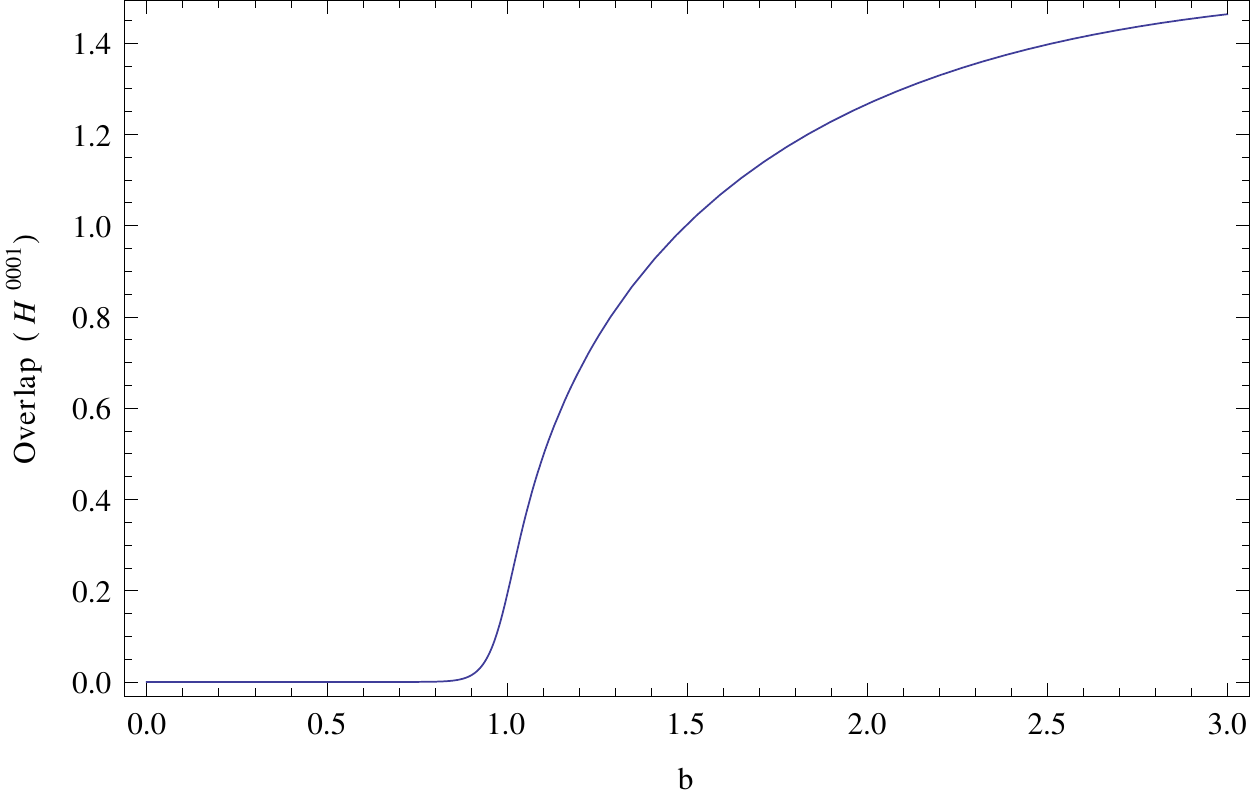}
  \end{tabular}
\caption{Figure shows the overlap integral $H^{0001}$ as a function of $b$ parameter of the scalar field.}
  \label{o1}
  \end{figure}
where we have defined another overlap integral $H^{mnlp}_{ijkl}$ as
\begin{equation}
 H^{mnlp}_{abcd}=\frac{1}{(\pi R)^\frac{a+b+c+d}{2}}\int dy~f^{(m)}_{H_a}f^{(n)}_{H_b}f^{(l)}_{H_c}f^{(p)}_{H_d}
\label{o1a}
 \end{equation}
while $K^{(m,n)}_{ab}$ and $G^{(m,n)}_{ab}$ are defined in Eq.(\ref{overlap1}). Here, $a,b,c,d=1,2$.
$f^{(m)}_{H_a}(y)$ is the bulk profile for the $m^{th}$ KK state of $H_a$.
Defining the mixing angle to be 
\begin{equation}
\delta=\frac{\mathcal{O}(v^2)H^{abcd}_{ijkl}}{M_{KK}^2} 
\label{mixingangle}
\end{equation}
 the mass eigenstate for the zero mode is now given as
\begin{equation}
 \phi^{(0)+}_{1M}=\phi^{(0)+}_{1M}-\delta\phi^{(1)+}_{1}
\end{equation}
Thus in the mass basis of the fermions, the coupling of two fermions to $\phi^{(0)+}_{1M}$ lead to additional sources of FCNC due to charged Higgs at the tree level.
Similar conclusions apply to the mixing of the zero mode of the real and the imaginary components of $H_a$ with their respective KK counterparts leading to neutral
FCNC at the tree level due to exchange of $h,H,A$. The, ``non-universality" in this case is parameterized by the overlap integral $H^{abcd}_{ijkl}$ defined in Eq.(\ref{o1a}).
As can be seen from Fig[\ref{o1}], this quantity is at-most $\mathcal{O}(1)$ for a scalar field localized near the IR brane ($b>1$) while it is negligible for a scalar
field localized near the UV brane ($b<1$).

\bibliographystyle{ieeetr}

       \bibliography{2HDM.bib}

\end{document}